\documentclass{article}

\usepackage{authblk}
\usepackage{graphicx,epsfig,color}
\usepackage{cite}
\usepackage[english]{babel}
\usepackage{amssymb,amsfonts,amsmath}
\usepackage{verbatim}
\usepackage{bbm,bbold}
\usepackage{bigints}
\newcommand{\cC}{{\cal C}}  
  \newcommand{\cF}{{\cal F}}
  
  \newcommand{\cJ}{{\cal J}}
  
\newcommand{\cM}{{\cal M}}  \newcommand{\cN}{{\cal N}}
\newcommand{\cO}{{\cal O}}  \newcommand{\cP}{{\cal P}}

\newcommand{\cW}{{\cal W}}

\newcommand{\be}{\begin{equation}} \newcommand{\ee}{\end{equation}}
\newcommand{\bea}{\begin{eqnarray}} \newcommand{\eea}{\end{eqnarray}}
\newcommand{\beann}{\begin{eqnarray*}}  \newcommand{\eeann}{\end{eqnarray*}}
\newcommand{\bfig}{\begin{figure}} \newcommand{\efig}{\end{figure}}
\newcommand{\ba}{\begin{array}} \newcommand{\ea}{\end{array}}
\newcommand{\bcen}{\begin{center}} \newcommand{\ecen}{\end{center}}
\newcommand{\btab}{\begin{tabular}} \newcommand{\etab}{\end{tabular}}

\newcommand{\vev}[1]{\left\langle{#1}\right\rangle}

\newtheorem{Proposition}{Proposition}[section]

\newtheorem{Theorem}{Theorem}[section]
\newtheorem{Lemma}{Lemma}[section]
\newtheorem{Corrolary}{Corrolary}[section]

\newcommand{\bp}{\begin{Proposition}}   \newcommand{\ep}{\end{Proposition}}
\newcommand{\bt}{\begin{Theorem}}   \newcommand{\et}{\end{Theorem}}
\newcommand{\bl}{\begin{Lemma}}     \newcommand{\el}{\end{Lemma}}
\newcommand{\bc}{\begin{Corrolary}} \newcommand{\ec}{\end{Corrolary}}
\newcommand{\JCS}[1]{{#1}}
\newcommand{\rr}[1]{{#1}}

\title{Effective long distance $q\bar{q} $ potential  in holographic RG flows}

\author[1]{Jorge Casalderrey-Solana} 
\author[2]{Diego Gutiez} 
\author[2]{Carlos Hoyos} 
\affil[1]{\small Departament de F\'\i sica Qu\`antica i Astrof\'\i sica \&  Institut de Ci\`encies del Cosmos (ICC), Universitat de Barcelona, Mart\'{\i}  i Franqu\`es 1, 08028 Barcelona, Spain}
\affil[2]{\small Department of Physics, Universidad de Oviedo, c/ Federico Garc\'ia Lorca, 18, 33007, Oviedo, Spain}

\begin{document}

\maketitle

\begin{abstract}
We study the $q\bar{q}$ potential in strongly coupled non-conformal field theories with a non-trivial renormalization group flow via holography. 
We focus on the properties of this potential at an inter-quark separation $L$ large compared to the characteristic scale of the field theory. These are determined by the 
leading order IR physics plus a series of corrections, sensitive to the properties of the RG-flow. To determine those corrections, we propose a general method applying holographic Wilsonian renormalization to a dual string. We apply this method to  examine in detail two sets of examples,  $3+1$-dimensional theories with 
%an RG flow from an IR fixed point driven by an irrelevant deformation; 
\rr{an RG flow ending in an IR fixed point;} and theories that are confining in the IR, in particular,  the Witten QCD and Klebanov-Strassler models. In both cases, we find corrections with a universal dependence on the inter-quark separation. When there is an IR fixed point, that correction decays as a power $\sim 1/L^4$. We  explain that dependence in terms of a double-trace deformation in a one-dimensional defect theory. For a confining theory, the decay is exponential $\sim e^{-ML}$, with $M$ a scale of the order of the glueball mass. 
 We interpret this correction using an effective flux tube description as produced by a background internal mode excitation induced by sources localized at the endpoints of the flux tube. 
  We discuss how these results could be confronted with lattice QCD data to test whether the description of confinement via the gauge/gravity is qualitatively correct.
\end{abstract}
\newpage

\tableofcontents

%%%%%%%%%%%%%%%%%%%%%%%%%%%%%%%%%%%%%%
\section{Introduction}
%%%%%%%%%%%%%%%%%%%%%%%%%%%%%%%%%%%%%%

The gauge/gravity correspondence \cite{Maldacena:1997re,Gubser:1998bc,Witten:1998qj}, or holographic duality, has been used extensively as a phenomenological tool to describe properties of strongly coupled systems in QCD and condensed matter (see \cite{Adams:2012th,DeWolfe:2013cua,Brambilla:2014jmp,Ammon:2015wua,2017stmc.book.....N,CasalderreySolana:2011us} for reviews on the topic). In most cases the gravity dual corresponds to a theory which is microscopically different from the actual system of interest, but whose properties at low energy/large distance compared to some characteristic scale can be qualitatively similar. 
\JCS{Since many relevant observables are mostly sensitive to the long distance physics, dissimilarities at high energy/small distances are frequently inconsequential. In}
the holographic dual description this means that only some part of the geometry is of relevance \JCS{for those observables}. More precisely,
\JCS{while } the geometry has an asymptotic boundary that is identified with the ultraviolet (UV) of the field theory, 
\JCS{infrared (IR) dynamics are controlled by the deep interior of the geometry and long distance observables are mostly sensitive to this region.}
 The problem of restricting the holographic description to the low energy effective theory has been approached multiple times in different contexts, e.g.~\cite{Bhattacharyya:2008jc,Iqbal:2008by,Eling:2009sj,Faulkner:2009wj,Bredberg:2010ky,Nickel:2010pr,Heemskerk:2010hk,Faulkner:2010jy,Charmousis:2010zz,Donos:2015gia}. 

In the holographic ``Wilsonian'' renormalization group (RG) flow approach of  \cite{Heemskerk:2010hk,Faulkner:2010jy} the effective IR description is obtained by introducing a cutoff in the holographic radial coordinate and ``integrating out'' degrees of freedom between the asymptotic boundary and the cutoff. This results in a description consisting of the dual theory below the cutoff plus a boundary action that determines the boundary conditions of the fields at the cutoff. The RG flow equations are obtained from the condition that the full on-shell action has to be independent of the cutoff. The boundary action is a functional of the values of the field at the cutoff, and, by mapping these values to operators in the field theory dual, it is interpreted as introducing multitrace deformations in the effective theory at the cutoff scale. The RG flow equations then become equations for the multi-trace couplings.

A particularly significant set of observables in gauge theories for which an IR effective description would be useful are Wilson loops. Their holographic dual description is a Nambu-Goto string with endpoints attached to the asymptotic boundary of the dual geometry  \cite{Maldacena:1998im,Rey:1998ik}. Even though this is a fairly simple setup, its holographic Wilsonian RG flow has not been worked out.\footnote{In \cite{Kiritsis:2014kua} it was suggested that the holographic RG flow will be given by a mean-curvature RG flow of the type described in \cite{Bakas:2007tm}. Although we do not discard that a map to a description of this form might exist, our results seem to correspond to a different type of flow.} We will partially fill in the blank by studying the expectation value of a Wilson loop corresponding to two static sources separated a fixed distance, much larger than the characteristic scale of the theory. 
This is equivalent to computing the quark-antiquark potential. 
\JCS{In this configuration, most of the profile of the dual string remains in the deep interior of the geometry and is mostly sensitive to IR physics. This observation will allow us to use the inter-quark distance $L$ as an expansion parameter to approximate the string profile and its energy. }
Applying the holographic Wilsonian RG flow approach, we will derive the effective IR behavior \JCS{of the potential and determine its most relevant long distance corrections}.

\JCS{We will use this method to analyze two different types of holographic constructions with well understood IR geometries}. 
The first type is dual to a strongly coupled field theory that possesses an IR fixed point, 
%with a flow triggered by an irrelevant scalar operator;
\rr{such that at higher energy scales the theory flows away from the IR fixed point in a way determined by an irrelevant scalar deformation.}
 The second type is dual to a confining theory. In the first case, the characteristic scale  appears in the coefficient of the irrelevant operator. At energies much lower than this scale, the theory is very close to being conformal, with only small corrections induced by the flow. We will introduce a cutoff much below this scale, in such a way that the dual geometry is close to an $AdS$ dual to the IR fixed point. In the case of a confining theory, the cutoff will be introduced at a scale much larger than the mass gap of the theory. For the confining theory we will use duals with explicit string theory constructions, the Witten QCD (WQCD) \cite{Witten:1998zw} and Klebanov-Strassler (KS) \cite{Klebanov:2000hb} models. \rr{For the case of an IR fixed point 
%there are no ten-dimensional geometries readily available, although 
there are several examples in five-dimensional supergravity \cite{Girardello:1998pd,Distler:1998gb,Khavaev:1998fb,Freedman:1999gp,Behrndt:1999ay,Khavaev:2000gb}. In principle these can be lifted to ten dimensions using general reduction formulas \cite{Khavaev:1998fb,Lu:1999bw,Cvetic:1999xp,Nastase:2000tu,Pilch:2000ue,Cvetic:2000nc,Lee:2014mla,Ciceri:2014wya,Baguet:2015sma}, but the examples where this has been carried out explicitly (e.g.\cite{Pilch:2000fu}) are even scarcer and the geometry turns out to be more complicated than what we will be considering in this work.} \JCS{In both those two families of models, the effective field theory approach we develop will allow us to determine generic properties of the potential, independent of the details of the UV behavior of the dual field theories. }

For a flow with an IR fixed point, we observe two types of corrections. One is produced by the non-trivial RG flow, and depends on the dimension of the leading irrelevant operator that drives the flow away from the \JCS{long distance conformal field theory (CFT)}. The other correction has a universal form, in the sense that it is independent of the dimensions of the leading irrelevant operator. We propose an interpretation in terms of an effective IR defect theory localized on the Wilson loop. In addition to the bulk RG flow, there is an RG flow on the defect triggered by a double trace deformation, that we identify from the boundary conditions of the string at the cutoff. \JCS{The $L$-dependence of both kinds of corrections depends solely on the properties of the IR field theory, and all the information about its UV structure is restricted to the value of a single coefficient, which determines the universal contribution.}

\JCS{For a flow in a confining theory we observe exponentially suppressed corrections to the potential beyond the leading linear dependence on the quark-antiquark separation}. The exponent is proportional to the mass scale of glueballs, and it coincides with the mass of some internal excitations of a flux tube in the dual field theory. We interpret this result in terms of the effective IR theory as a flux tube with sources at the endpoints for the internal modes, \JCS{ that have a non-vanishing profile in the ground state of the quark-antiquark pair}. 
\JCS{Since these excitations correspond to fluctuations of the string along the holographic direction, these type of correction are a generic property of confinement as  described by the gauge/gravity duality.}

The structure of the paper is as follows: we start reviewing some basic facts about Wilson loops and their calculation using gauge/gravity duality in \S~\ref{sec:review}. In \S~\ref{sec:rgflow} we present general formulas for the holographic RG flow of a Wilson loop. The case with an IR fixed point is studied in \S~\ref{sec:ircft} and confining theories in \S~\ref{sec:conf}. A summary  of the results and their interpretation from the point of view of the field theory dual is gathered in \S~\ref{sec:discuss}.

%%%%%%%%%%%%%%%%%%%%%%%%%%%%%%%%%%%%%%
\section{Wilson loops in holography}\label{sec:review}
%%%%%%%%%%%%%%%%%%%%%%%%%%%%%%%%%%%%%%

The study of Wilson loops in holographic duals was initiated in \cite{Maldacena:1998im,Rey:1998ik}. In $\cN=4$ super Yang-Mills (SYM), a locally BPS Wilson loop in the fundamental representation is given by the path-ordered exponential
\be
\cW_{BPS}(\cC)=\frac{1}{N}{\rm Tr}\,\cP\, \left( e^{i\oint_\cC d\tau\left( \dot{x}^\mu A_\mu +|\dot{x}| \theta_I \Phi^I\right)}\right),\ \ \theta^2=1.
\ee
Where  $x^\mu(\tau)$ parametrizes the closed curve $\cC$ on which the loop is defined, $A_\mu$ is the gauge field and $\Phi^I$, $I=1,\cdots,6$ are the adjoint scalar fields of the $\cN=4$ SYM theory. In the large-$N$ limit the expectation value of the Wilson loop can be determined at strong 't Hooft coupling $\lambda_{YM}\gg1 $ by the classical Nambu-Goto action of an open string whose boundary is along the curve $\cC$ at the asymptotic boundary of the holographic dual geometry. The position of the string endpoints in the internal space is determined by identifying the functions $\theta^I(\tau)$ with coordinates on the $S^5$ of the dual $AdS_5\times S^5$ geometry. 

The identification of BPS loops with a dual string is based on the weakly coupled D-brane construction. The $\cN=4$ SYM theory is the low energy description of a stack of coincident  $N$ D3 branes. When one of the branes is separated from the rest, a string extended between the separated brane and the rest acts as a source in the fundamental representation. When the isolated brane is taken to infinity, the string becomes an infinitely heavy source and therefore it is equivalent to the insertion of a Wilson loop. In the near-horizon limit that replaces D3 branes by geometry, the isolated brane can be though as being at the asymptotic boundary, and the string extends from it to the interior. A similar argument can be used for any low energy effective theory on the worldvolume of a stack of branes, so the identification of the string with a Wilson loop extends naturally to more general gauge/gravity duals obtained from a near-horizon limit.

We will see that the holographic Wilsonian RG flow changes the boundary conditions of the string at the cutoff. This could alter the nature of the Wilson loop, for instance it was proposed in \cite{Alday:2007he} that the holographic dual to an ordinary Wilson loop should correspond to a string satisfying Neumann boundary conditions along the $S^5$ directions. The logic is that an ordinary Wilson loop does not couple to the scalars $\Phi^I$, and therefore it is invariant under the $SO(6)$ R-symmetry that rotates them. More generally, one could define a family of Wilson loops with different couplings to the scalar fields  \cite{Polchinski:2011im}
\be
\cW_{\zeta}(\cC)=\frac{1}{N}{\rm Tr}\,\cP\, \left( e^{i\oint_\cC d\tau\left( \dot{x}^\mu A_\mu +\zeta|\dot{x}| \theta_I \Phi^I\right)}\right),
\ee
with $\zeta=1$ corresponding to the BPS loop and $\zeta=0$ to the ordinary Wilson loop. In  \cite{Polchinski:2011im} it was shown that there can be an RG flow between the ordinary and BPS Wilson loops, both at weak and strong coupling.  For Wilson loops with intermediate values of $\zeta$ one may expect that the dual description is a string with mixed boundary conditions. The case at hand differs from the $\zeta$-deformed loops in that the boundary conditions that are modified are not along the $S^5$, but along the field theory directions. There are however some similarities, in that we can identify a deformation of the BPS loop and an associated RG flow.

%%%%%%%%%%%%%%%%%%%%%%%%%%%%%%%%%%%%%%
\subsection{Calculation of the quark-antiquark potential in an RG flow}
%%%%%%%%%%%%%%%%%%%%%%%%%%%%%%%%%%%%%%

The holographic dual to a conformal field theory is an $AdS_5\times \cM_5$ geometry, where $\cM_5$ is a compact space. In Gaussian coordinates
\be\label{eq:adsspace}
ds_{10}^2=G_{MN}dx^M dx^N=dr^2+e^{2r/R}\eta_{\mu\nu}dx^\mu dx^\nu+d\cM_5^2,
\ee
where $x^\mu=(t,x,y,z)$ are coordinates along the field theory directions and $\eta_{\mu\nu}$ is the flat Minkowski metric. The coordinates along the compact space will be denoted by $\theta^A$, $A=1,\cdots,5$. The radial coordinate $r$ characterizes the energy/distance scale in the field theory, with $r\to \infty$ the asymptotic boundary associated to the UV. $R$ is the radius of $AdS$ space. 

The potential $V_{q\bar{q}}$ between a static quark-antiquark pair separated a distance $L$ can be computed from the expectation value of a time-like Wilson loop along a rectangular contour of sides of length $L$ along space and $\beta$ along time. When $\beta\to \infty$,
\be
\vev{\cW}\sim e^{-\beta V_{q\bar{q}}(L)}.
\ee
In the large-$N$ limit, and at strong 't Hooft coupling, the potential is determined by the Nambu-Goto action of a string evaluated on-shell
\be
\beta\, V_{q\bar{q}}(L)=-S_{NG}.
\ee 
The Nambu-Goto action is
\be
S_{NG}=-\frac{1}{2\pi\alpha'}\int d^2 \sigma \sqrt{-h}, \ \ h_{ab}=G_{MN}\partial_a X^M \partial_b X^N, \ \ \sigma^a=(\tau,\sigma),
\ee
where $\sigma^a$ are the world-sheet coordinates, $h$ is the determinant of the induced metric $h_{ab}$ and $X^M(\tau,\sigma)$ are the embedding functions that describe the string profile in the target space.

The relevant configuration is a solution to the classical equations of motion with appropriate boundary conditions, \rr{following previous works (e.g.\cite{Kinar:1998vq}) we review here the main points of the derivation and properties of the solutions.} We can choose the following static gauge
\be\label{eq:stringemb}
X^0=\tau,\ \ X^1=x(\sigma),\ \ X^2=X^3=0,\ \ X^r=\sigma,\ \ X^A=\theta^A_0,
\ee
where $\theta^A_0$ are constant. The boundary conditions are
\be\label{eq:boundarycond}
\lim_{\sigma\to \infty} x(\sigma)=0, \ \ \lim_{\sigma\to \sigma_*} x'(\sigma)=\infty, \ \  \lim_{\sigma\to \sigma_*} x(\sigma)=\frac{L}{2}.
\ee
\JCS{where $\sigma_*$ is a particular value of the world-sheet coordinate $\sigma$, whose value depends on $L$, via the last condition.}
This solution actually describes a branch of the solution extending from the asymptotic boundary to a point in the interior, there is another symmetric branch returning to the boundary at the point $x=L$.

In general the action evaluated on this class of solutions is divergent, one can regularize it by introducing a cutoff at $\sigma=r_{(\Lambda)}$ and adding a counter-term at the boundary of the string, \rr{in such a way that the total action is $S_{\rm string}=S_{NG}+S_{c.t.}$}, so that
\be\label{eq:stringaction}
S_{\rm string}=\lim_{r_{(\Lambda)}\to\infty}\;-\frac{1}{2\pi\alpha'}\int_{\sigma\leq r_{(\Lambda)}} d^2 \sigma \sqrt{-h}+\frac{c_\Lambda}{2\pi \alpha'}\int d\tau \,\sqrt{-\gamma}, %\ \gamma=h_{\tau\tau}\Big|_{\sigma=r_{(\Lambda)}}.
\ee
\rr{where $\gamma=h_{\tau\tau}\Big|_{\sigma=r_{(\Lambda)}}$.} For an asymptotically $AdS$ space like \eqref{eq:adsspace}, the value of the coefficient of the counter-term is $c_\Lambda=R$.

The holographic dual of a generic RG flow can be a relatively complicated geometry with various warping factors depending on the radial coordinate and coordinates along the internal space 
\be
ds_{10}^2=\Delta(\theta,r)dr^2+\Sigma(\theta,r)\eta_{\mu\nu}dx^\mu dx^\nu+d\widetilde{\cM}_5^2.
\ee
\JCS{In this work, we will concentrate in simpler examples, in which the ten-dimensional metric can be put in the domain wall form}
\be\label{eq:domainwall}
ds_{10}^2=\frac{dr^2}{f(r)}+e^{2A(r)}\eta_{\mu\nu}dx^\mu dx^\nu+d\widetilde{\cM}_5^2.
\ee
Where, as the asymptotic boundary at $r\to \infty$ is approached, $A(r)\to \infty$ and $f(r)\to 1$.

In this case \eqref{eq:stringemb}  is a consistent ansatz and the induced metric is
\be
ds_2^2=-e^{2A(\sigma)}d\tau^2 + \left[\frac{1}{f(\sigma)}+e^{2A(\sigma)}(x')^2\right]d\sigma^2.
\ee
The action becomes
\be
S_{NG}=-\frac{\beta}{2\pi\alpha'}\int d\sigma \, \frac{e^A}{\sqrt{f}} \sqrt{1+f e^{2A}(x')^2}.
\ee
Again, one needs to add a counter-term of the form given in \eqref{eq:stringaction} to render it finite.

\begin{figure}[t!]
\begin{center}
\includegraphics[width=10cm]{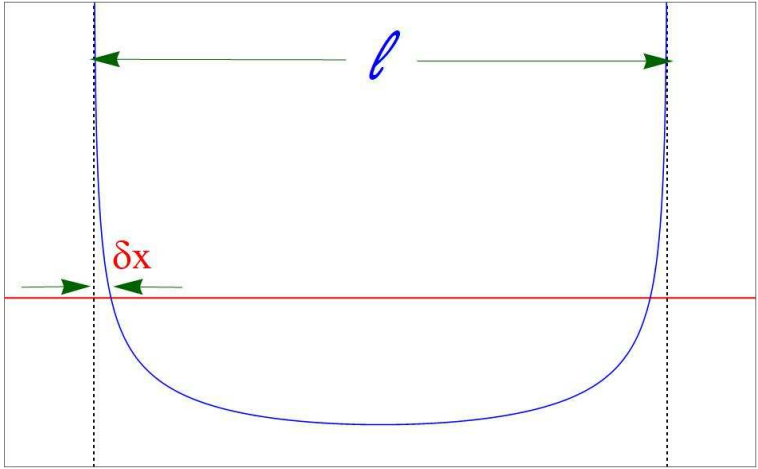}
\end{center}
\caption{Profile of a string dual to a $q\bar{q}$ pair separated a distance $\ell$ (blue line). The vertical direction corresponds to the holographic radial coordinate, with the asymptotic boundary (UV) at the top. A cutoff is introduced at an IR scale (horizontal red line) and degrees of freedom above the cutoff are integrated out. The separation in the field theory directions between the endpoint of the string at the boundary and at the cutoff is denoted by $\delta x$, it corresponds to $x_{(\mu)}$ in the text. }\label{fig:wl}
\end{figure}

Since it depends only on the derivative of the embedding function $x'$, the conjugate momentum is constant 
\be
\pi_x=\frac{\delta S_{NG}}{\delta x'}=\frac{\beta}{2\pi\alpha'} p={\rm constant}.
\ee
This leads to the equation
\be\label{eq:eom}
\frac{\sqrt{f} e^{3A} x'}{\sqrt{1+f e^{2A}(x')^2}}=-p,
\ee
or, solving for $x'$, we obtain the equation of motion for the embedding
\be\label{eq:xsol}
x'=-p\frac{e^{-3A}}{\sqrt{f}\sqrt{1-e^{-4A}p^2}}.
\ee
If we picture the string as hanging from the asymptotic boundary, as in Fig.~\ref{fig:wl}, the conditions \eqref{eq:boundarycond} fix the relation between the lowest point of the string profile $\sigma_*$, the separation $L$ of the pair and $p$
\be\label{eq:pval}
p=e^{2 A(\sigma_*)},\ \ \frac{L}{2}=p \int_{\sigma_*}^\infty \frac{e^{-3A}}{\sqrt{f}\sqrt{1-e^{-4A}p^2}}.
\ee
Note that under a change of the boundary condition $\delta x(\infty)=\delta L$, the solution changes $x\to x+\delta x$ and the change of the Nambu-Goto action is proportional to the conjugate momentum
\be
\delta S_{NG}=\int^\infty  d\sigma \, \pi_x \delta x'=\frac{\beta}{2\pi\alpha'} p \,\delta L.
\ee
The change of the potential is
\be
\delta V_{q\bar{q}}=-\frac{1}{2\pi\alpha'} p \delta L. 
\ee
Therefore, $p$ should be identified with the force that the (anti)quark feels 
\be\label{eq:force}
\cF_x=-\frac{\delta V_{q\bar{q}}}{\delta L}=\frac{1}{2\pi\alpha'} p.
\ee

%%%%%%%%%%%%%%%%%%%%%%%%%%%%%%%%%%%%%%
\section{Holographic RG flow of the Wilson loop}\label{sec:rgflow}
%%%%%%%%%%%%%%%%%%%%%%%%%%%%%%%%%%%%%%

\JCS{The analysis of the previous section demanded to have full knowledge of the dual geometry all the way to the boundary to determine the classical string profile that controls the potential. However, for large enough separation it is expected that it is enough to know the geometry below certain cutoff}  $r_{(\mu)}$. \JCS{To materialize this expectation, we note that for sufficiently large values of $L$,  the profile of the string is typically mostly below some finite value of the radial cutoff $r_{(\mu)}$}.
In the dual geometry the region between the boundary and the cutoff corresponds to length scales much smaller than  the separation $L$. At these scales, the quark and antiquark do not feel much the presence of each other, and the profile of the string around each of the endpoint positions is close to that of a single isolated quark, remaining close to the endpoint position in the parallel directions to the boundary and extending almost completely straight into the interior. The straight shape of the profile persists from the cutoff to the interior, until, far from the cutoff, the profile changes and extends in the directions parallel to the boundary in such a way that the two endpoints are joined (see Figure~\ref{fig:wl}). This characteristic behavior implies that the information about the UV properties of the theory is confined to the position of the string profile at the cutoff, which is close to the position of the endpoints at the boundary and introduces a length scale much smaller than the separation between the endpoints. The ratio between these two length scales works as a perturbative parameter that we will use to find the first corrections to the leading order dependence of the potential on the quark-antiquark separation.

Based on the considerations above, we separate the contribution of the straight segments from the rest by introducing a cutoff at $\sigma=r_{(\mu)}$ such that
\be
x_{(\mu)} =x(r_{(\mu)})\ll L. 
\ee
In the region closer to the asymptotic boundary $\sigma>r_{(\mu)}$, the string does not wander far from its initial position $0<x< x_{(\mu)}$, and $x'$ is a small quantity, as shown in Fig.~\ref{fig:wl}. Taking advantage of this fact, we will split the string action in two parts, an upper part \rr{ $S_{\rm string}^>$}, where we integrate for values $\sigma>r_{(\mu)}$ \rr{ and add the boundary counterterms}, and the lower part where we integrate below the cutoff \rr{$S_{NG}^<$}.
\be
S_{\rm string}=S_{\rm string}^>+S_{\rm NG}^<=S_{NG}^>+S_{c.t.}+S_{NG}^<.
\ee
Expanding the upper action to quadratic order one finds
\be
S_{NG}^>\simeq -\frac{\beta}{2\pi \alpha'}\int_{\sigma>r_{(\mu)}} d\sigma \frac{e^A}{\sqrt{f}}\left( 1+\frac{1}{2}fe^{2A}(x')^2\right).
\ee
The solution to the equations of motion is
\be
x'\simeq -p \frac{e^{-3A}}{\sqrt{f}}.
\ee
Comparing with the exact solution \eqref{eq:xsol},  this is a valid approximation as long as $p^2 e^{-4A(r_{(\mu)})}\ll 1$.  Defining a function
\be\label{eq:amu}
a(\sigma)=\int_{\sigma}^\infty \frac{e^{-3A}}{\sqrt{f}}, \ \ a_{(\mu)}=a(r_{(\mu)}),
\ee
the displacement at the cutoff is $x_{(\mu)}= a_{(\mu)} p$. The profile of the string is then
\be
x\simeq x_{(\mu)}\frac{a(\sigma)}{a_{(\mu)}}.
\ee
The on-shell action (regularized by an UV cutoff) is
\be
S_{NG}^>\simeq -\frac{\beta}{2\pi \alpha'}\left[\int_{r_{(\mu)}}^{r_{(\Lambda)}} d\sigma \frac{e^A}{\sqrt{f}}+\frac{\sqrt{f}e^{3A}}{2}x' x\Big|_{r_{(\mu)}}^{r_{(\Lambda)}}\right].
\ee
The string action, including counterterms, is thus
\be\label{eq:boundact}
S_{\rm string}^>\simeq -\frac{\beta}{2\pi \alpha'}\left[ M_{(\mu)}+\frac{1}{2a_{(\mu)}}  x_{(\mu)}^2\right],
\ee
where
\be\label{eq:Mmu}
M_{(\mu)}=\lim_{r_{(\Lambda)}\to\infty}\int_{r_{(\mu)}}^{r_{(\Lambda)}} d\sigma \frac{e^A}{\sqrt{f}}-e^{A(r_{(\Lambda)})},\\
\ee
We can then replace our original string action by the NG action below the cutoff plus a boundary term that appears as a double-trace deformation, the $x_{\mu}^2$ term appearing in \eqref{eq:boundact}, its effect is to modify the boundary conditions at the cutoff. In order to see this, consider a string with slightly perturbed profile, but keeping the endpoints at the boundary fixed $x\to x+\delta x$, $\delta x(r_{(\Lambda)})=0$. The variation of the on-shell string action under this perturbation has a bulk contribution and a contribution localized at the cutoff
\be
\delta S_{\rm string}=-\frac{\beta}{2\pi\alpha'}\left[\frac{x_{(\mu)}}{a_{(\mu)}}\delta x_{(\mu)} +\int_{\sigma< r_{(\mu)}} d\sigma\, \frac{\sqrt{f}e^{3A}x'}{\sqrt{1+f e^{2A}(x')^2}}\delta x'\right].
\ee
Integrating the bulk term by parts and using the equations of motion \eqref{eq:eom}, one is left with only cutoff contributions
\be
\delta S_{\rm string}=-\frac{\beta}{2\pi\alpha'}\left[\frac{x_{(\mu)}}{a_{(\mu)}}-p\right]\delta x_{(\mu)} .
\ee
Since the on-shell action should be stationary for small perturbations of the profile that do not change the boundary conditions, the variation above should vanish for any $\delta x_{(\mu)}$. This condition fixes the conjugate momentum for the solution below the cutoff to the right value
\be
p=\frac{x_{(\mu)}}{a_{(\mu)}}.
\ee
Therefore, the string dual to the Wilson loop that determines the quark-antiquark potential can be replaced by a string with endpoints at a cutoff satisfying mixed boundary conditions. 

%%%%%%%%%%%%%%%%%%%%%%%%%%%%%%%%%%%%%%
\subsection{IR description of the Wilson loop}
%%%%%%%%%%%%%%%%%%%%%%%%%%%%%%%%%%%%%%
\JCS{The analysis above has made precise the expectation that the long distance potential only depends on the IR physics. By replacing the full string action from the cut-off to the boundary by the quadratic approximation, \eqref{eq:boundact}, we have managed to express the problem in terms of quantities evaluated at the cut-off. All the information about the UV part of the geometry, and its manifestation in the string embedding, is condensed into the (cut-off dependent) values of the parameters $M_{(\mu)}$ and $a_{(\mu)}$. Starting from this action, in this section we will show how to use the independence of physical quantities on the cutoff to constraint the long distance behavior of the heavy quark potential. }

Suppose we are given a geometry that will be used as a holographic dual description of the IR physics of some strongly coupled theory. We introduce a cutoff in this geometry and consider the string action with the additional boundary terms we have derived
\be
S_{IR}=S_{NG}^<-\frac{\beta}{2\pi \alpha'}\left[ M_{(\mu)}+\frac{1}{2a_{(\mu)}}  x_{(\mu)}^2\right].
\ee
Physical quantities computed using the holographic dual should be independent of the cutoff we have introduced. However, the string action has an explicit dependence on the cutoff, that is apparent from the definition of the coefficients of the cutoff terms $a_{(\mu)}$ \eqref{eq:amu} and $M_{(\mu)}$ \eqref{eq:Mmu}. If we regard $r_{(\mu)}$ as corresponding to an RG scale similar to the ones used in perturbative renormalization schemes, the dependence on the cutoff can be encoded in the RG flow equations ($A_{(\mu)}=A(r_{(\mu)})$, $f_{(\mu)}=f(r_{(\mu)})$)
\be\label{eq:rgflowam}
\partial_{r_{(\mu)}}a_{(\mu)}= - \frac{e^{-3A_{(\mu)}}}{\sqrt{f_{(\mu)}}},\ \ \partial_{r_{(\mu)}}M_\mu=-\frac{e^{A_{(\mu)}}}{\sqrt{f_{(\mu)}}}.
\ee
Integrating these equations one would obtain $M_{(\mu)}$ and $a_{(\mu)}$ up to indeterminate integration constants. It should be noted that the RG equations involve terms that are evaluated at the cutoff position, so they only depend on the local geometry close to the cutoff. In the language of the field theory dual, the RG equations only depend on the physics of the scale close to the cutoff. All the information about UV physics is hidden in the integration constants. This fits with the usual Wilsonian paradigm of renormalisation, the terms that can appear in the effective action are determined by the IR degrees of freedom, but with coefficients that have to be fixed by experiments or by matching with UV physics.

Using the equations \eqref{eq:rgflowam} it is easy to show that physical quantities are independent of the cutoff.  The equations of motion for the embedding below the cutoff are given by \eqref{eq:xsol}, and we have to impose the conditions
\be\label{eq:stringcond}
p=e^{2 A(\sigma_*)},\ \ x_{(\mu)}=a_{(\mu)} p,\ \ \frac{L}{2}=x_{(\mu)}+p \int_{\sigma_*}^{r_{(\mu)}}\frac{e^{-3A}}{\sqrt{f}\sqrt{1-e^{-4A}p^2}}.
\ee
The force, which is proportional to $p$, is independent of the cutoff by construction. The separation between the quark-antiquark pair is invariant under changes of the cutoff at leading order
\be
\partial_{r_{(\mu)}}L=2p\left( \partial_{r_{(\mu)}}a_{(\mu)}+\frac{e^{-3A_{(\mu)}}}{\sqrt{f_{(\mu)}}\sqrt{1-e^{-4A_{(\mu)}}p^2}}\right)=O\left(p^3 e^{-7A_{(\mu)}}\right).
\ee
Then, the dependence of the force \eqref{eq:force} with the length is also invariant to leading order. In principle one could systematically add higher order corrections by including further multi-trace terms in the boundary action, \rr{we show how to proceed using the Wilsonian RG flow equations for the boundary action and show explicitly that the length is invariant at the next order in Appendix~\ref{sec:Sflow}}. As we already mentioned, the value of $a_{(\mu)}$ cannot be determined by the IR theory, rather it would have to be fixed by matching with the UV theory if this one is known, or by measuring the force at a separation $L$ and doing a fit.

%%%%%%%%%%%%%%%%%%%%%%%%%%%%%%%%%%%%%%
\section{Theory with an IR fixed point}\label{sec:ircft}
%%%%%%%%%%%%%%%%%%%%%%%%%%%%%%%%%%%%%%
\JCS{In this section we will use the formalism developed in the previous section to study a particularly simple example, that of a strongly coupled field theory with an IR fixed point. Because it is a fixed point, the long 
distance dynamics are controlled by a CFT. As a consequence of conformal symmetry, the dual geometry must approach $AdS$ space in the interior,}
meaning it takes the form in \eqref{eq:adsspace} as $r\to-\infty$. 
We will assume that the flow away from the fixed point is driven an irrelevant scalar operator of conformal dimension $\Delta>d$. That is, to the action of the IR CFT we add a coupling to the irrelevant operator
\be
S_{IR}=S_{CFT}+\int d^dx\,\tilde \alpha \cO_\Delta.
\ee
In the dual theory, this is realized by turning on a scalar field $\phi$ of mass $m^2R^2=\Delta(\Delta-d)$. The scalar field back-reacts on the geometry and takes it away from $AdS$ to a different geometry in the asymptotic region, which in principle could be another $AdS$ space of different radius $R_{UV}>R$, corresponding to a UV fixed point.

In order to simplify the discussion we will use the `fake' supergravity formalism \cite{Skenderis:1999mm,Freedman:2003ax,Celi:2004st,Zagermann:2004ac,Skenderis:2006jq},\footnote{\rr{The formalism can be derived from a Hamilton-Jacobi formulation of the radial evolution of the solutions, which has also been interpreted in terms of the RG flow in the holographic dual, see e.g.\cite{deBoer:1999tgo,Kiritsis:2016kog,Nitti:2017cbu} and \cite{Salopek:1990jq} for an earlier work in the context of cosmology.}} such that the equations of motion for the metric and the scalar reduce to a system of first order equations. It should be noted that for any given potential in supergravity, a superpotential describing the solution (fake or not) always exists locally, so the analysis presented here for the leading corrections is quite general. 

The metric  has a slightly simpler form than the general case
\be
ds_5^2=dr^2+e^{2A(r)}\eta_{\mu\nu}dx^\mu dx^\nu.
\ee
\JCS{Note, in particular, that the function $f(r)=1$, as defined in \eqref{eq:domainwall}.}
The equations of motion for the metric and scalar are
\be
\phi'=-\partial_\phi W, \ \ A'=\frac{W}{d-1}.
\ee
There is a critical point at a value $\phi=\phi_{IR}$ such that
\be
\partial_\phi W(\phi_{IR})=0,\ \ W(\phi_{IR})= \frac{d-1}{R}.
\ee
This corresponds to the IR $AdS$ solution where the scalar is constant
\be
\phi(r)=\phi_{IR},\ \ A(r)=\frac{r}{R}.
\ee
Close to the critical point, the fake superpotential can be approximated by
\be
W\simeq \frac{d-1}{R}+\frac{d-\Delta}{2R}(\phi-\phi_{IR})^2.
\ee
The solution to leading order \JCS{away from conformality} is
\be
\phi \simeq \phi_{IR}+\alpha e^{(\Delta-d)(r-r_{(M)})/R},\ \ A\simeq \frac{r}{R}-\frac{\alpha^2}{4}e^{2(\Delta-d)(r-r_{(M)})/R},
\ee
where $\alpha$ is of order one and proportional to the irrelevant coupling, $\tilde \alpha$, and $r_{(M)}$ determines the region where the geometry deviates significantly from $AdS$. The type of expansion we are doing is valid for $r\ll r_{(M)}$. In particular, the cutoff should be in the near-$AdS$ region $r_{(\mu)}\ll r_{(M)}$.

From the first condition in \eqref{eq:stringcond}, we obtain
\be
p\simeq \exp\left( \frac{2\sigma_*}{R}-\frac{\alpha^2}{2}e^{2(\Delta-d)(\sigma_*-r_{(M)})/R} \right).
\ee
We can solve this condition for $\sigma_*$ expanding to leading order in $\alpha$
\be
\sigma_*=\sigma_*(0)+\delta\sigma\simeq \frac{R}{4}\log p^2+R\frac{\alpha^2}{4}e^{-2(\Delta-d)r_{(M)}/R}p^{\Delta-d}.
\ee
From \eqref{eq:stringcond}, the quark-antiquark separation is
\be
L=2a_{(\mu)} p+2p \int_{\sigma_*}^{r_{(\mu)}} I(\sigma),
\ee
where
\be
I(\sigma)=\frac{e^{-3A}}{\sqrt{1-e^{-4A}p^2}}.
\ee
Let us split the integral in a region close to $\sigma_*$ and the rest
\be
\int_{\sigma_*}^{r_{(\mu)}} I(\sigma)=\int_{\sigma_*}^{\sigma_*+\Delta\sigma} I(\sigma)+\int_{\sigma_*+\Delta\sigma}^{r_{(\mu)}} I(\sigma).
\ee
The first integral is approximated expanding around $\sigma_*$. Expanding the result in $\alpha$, one finds
\be
\int_{\sigma_*}^{\sigma_*+\Delta\sigma} I(\sigma)\simeq \frac{R^{1/2}}{p^{3/2}}\left(1+(\Delta-d)\frac{\alpha^2}{4}e^{-2(\Delta-d)r_{(M)}/R}p^{\Delta-d} \right)\sqrt{\Delta\sigma}.
\ee
In the second integral we expand first in $\alpha$
\be
\begin{split}
\int_{\sigma_*+\Delta\sigma}^{r_{(\mu)}} I(\sigma)&\simeq -\frac{e^{-3\sigma/R}}{\sqrt{1-e^{-4\sigma/R}p^2}}\Big|_{\sigma_*(0)+\Delta\sigma}\delta\sigma\\
&+\int_{\sigma_*(0)+\Delta\sigma}^{r_{(\mu)}} \left[\frac{e^{-3\sigma/R}}{\sqrt{1-e^{-4\sigma/R}p^2}}+\frac{\alpha^2}{4}e^{-2(\Delta-d)r_{(M)}/R} \frac{e^{(2(\Delta-d)-3)\sigma/R}\left( 3-e^{-4\sigma/R}p^2\right)}{\left(1-e^{-4\sigma/R}p^2\right)^{3/2}} \right].
\end{split}
\ee
The integrals can be done analytically, with a result that can be expressed in terms of hypergeometric functions, but that is not very illuminating. Expanding for $r_{(\mu)}\gg \sigma_*(0)$, and $\Delta \sigma \to 0$,
\be
\frac{e^{-3\sigma/R}}{\sqrt{1-e^{-4\sigma/R}p^2}}\Big|_{\sigma_*(0)+\Delta\sigma}\delta\sigma\simeq \frac{\alpha^2}{4}e^{-2(\Delta-d)r_{(M)}/R}\frac{R^{3/2}}{2\sqrt{\Delta\sigma}}p^{\Delta-d-\frac{3}{2}},
\ee
\be
\int_{\sigma_*(0)+\Delta\sigma}^{r_{(\mu)}} \,\frac{e^{-3\sigma/R}}{\sqrt{1-e^{-4\sigma/R}p^2}}\simeq -\frac{R^{1/2}}{p^{3/2}}\sqrt{\Delta\sigma}+\frac{R}{p^{3/2}} \frac{\sqrt{\pi}\Gamma\left(\frac{3}{4}\right)}{\Gamma\left(\frac{1}{4}\right)}-\frac{R}{3}e^{-3r_{(\mu)}/R},
\ee
\be
\begin{split}
&\int_{\sigma_*(0)+\Delta\sigma}^{r_{(\mu)}} \frac{e^{(2(\Delta-d)-3)\sigma/R}\left( 3-e^{-4\sigma/R}p^2\right)}{\left(1-e^{-4\sigma/R}p^2\right)^{3/2}}\simeq \frac{R^{3/2}}{2\sqrt{\Delta\sigma}}p^{\Delta-d-\frac{3}{2}}-(\Delta-d+1) R^{1/2} p^{\Delta-d-\frac{3}{2}}\sqrt{\Delta\sigma}\\
&+\frac{3R}{2(\Delta-d)-3} e^{(2( \Delta-d)-3)r_{(\mu)}/R}+\frac{(\Delta-d)\sqrt{\pi}\Gamma\left( \frac{3}{4}-\frac{\Delta-d}{2}\right)}{2\Gamma\left(\frac{5}{4}-\frac{\Delta-d}{2}\right)}R p^{\Delta-d-\frac{3}{2}}.
\end{split}
\ee
The $\sim 1/\sqrt{\Delta\sigma}$ terms cancel out when we sum \JCS{over} all contributions, so the limit $\Delta\sigma\to 0$ is finite, giving
\be
\begin{split}
\int_{\sigma_*}^{r_{(\mu)}} I(\sigma)&\simeq R\left[ \frac{\sqrt{\pi}\Gamma\left(\frac{3}{4}\right)}{\Gamma\left(\frac{1}{4}\right)}p^{-3/2}-\frac{1}{3}e^{-3r_{(\mu)}/R}\left(1-\frac{9}{4}\frac{\alpha^2}{2(\Delta-d)-3} e^{-2(\Delta-d)(r_{(M)}-r_{(\mu)})/R}\right)\right.\\
&\left.+\frac{\alpha^2}{4}e^{-2(\Delta-d)r_{(M)}/R} \frac{(\Delta-d)\sqrt{\pi}\Gamma\left( \frac{3}{4}-\frac{\Delta-d}{2}\right)}{2\Gamma\left(\frac{5}{4}-\frac{\Delta-d}{2}\right)} p^{\Delta-d-\frac{3}{2}}\right].
\end{split}
\ee 
 The expression is valid for $\Delta-d\neq 3/2$. In order for the correction proportional to $\alpha^2$ to be small, the cutoff should be at a position in the radial direction $r_{(\mu)}\ll r_{(M)}$. In the field theory dual this means that we are considering energy scales much below the characteristic scale where the RG flow deviates significantly from the IR CFT. For $\Delta-d=3/2$ one finds
\be
\begin{split}
\int_{\sigma_*}^{r_{(\mu)}} I(\sigma)&\simeq R\left[ \frac{\sqrt{\pi}\Gamma\left(\frac{3}{4}\right)}{\Gamma\left(\frac{1}{4}\right)}p^{-3/2}-\frac{1}{3}e^{-3r_{(\mu)}/R}+\frac{\alpha^2}{4}e^{-3 r_{(M)}/R} \left( 3 \frac{r_{(\mu)}}{R}-\frac{3}{2}\log\left( \frac{p}{2}\right)-1\right)  \right].
\end{split}
\ee 

The quark-antiquark separation is
\be\label{eq:length1}
L=R\left[ c_0 p^{-1/2}+ a_0 p+a_{\Delta-d} p^{\Delta-d-\frac{1}{2}}\right], \ \ \Delta-d\neq \frac{3}{2},
\ee
or
\be\label{eq:length2}
L=R\left[ c_0 p^{-1/2}+ \tilde{a}_0 p+a_{3/2}  p\log \frac{p}{p_0}\right], \ \ \Delta-d=\frac{3}{2}.
\ee 
The coefficients are
\be
\begin{split}
&c_0=\frac{2\sqrt{\pi}\Gamma\left(\frac{3}{4}\right)}{\Gamma\left(\frac{1}{4}\right)},\ \ a_0=2a_{(\mu)}-\frac{2}{3}e^{-3r_{(\mu)}/R}\left(1-\frac{9}{4}\frac{\alpha^2}{2(\Delta-d)-3} e^{-2(\Delta-d)(r_{(M)}-r_{(\mu)})/R}\right),\\
&a_{\Delta-d}=\frac{\alpha^2}{2}e^{-2(\Delta-d)r_{(M)}/R} \frac{(\Delta-d)\sqrt{\pi}\Gamma\left( \frac{3}{4}-\frac{\Delta-d}{2}\right)}{2\Gamma\left(\frac{5}{4}-\frac{\Delta-d}{2}\right)},\\
&\tilde{a}_0=2 a_{(\mu)}-\frac{2}{3}e^{-3r_{(\mu)}/R}\left(1-\frac{9\alpha^2}{4}e^{-3 (r_{(M)}-r_{(\mu)})/R} \frac{r_{(\mu)}}{R}\right),\\
&a_{3/2}=-\frac{3\alpha^2}{4}e^{-3 r_{(M)}/R},\ \ p_0=2e^{-2/3}.
\end{split}
\ee
Only the coefficients $a_0$ and $\tilde{a}_0$ have \JCS{an explicit} dependence on the cutoff. 
\JCS{However, when taking into account the scale dependence of the double trace coefficient $a_{(\mu)}$, this dependence vanishes.}
Expanding the RG-flow equations for $a_{(\mu)}$ \eqref{eq:rgflowam} to $O(\alpha^2)$, one can show that
\be
\partial_{r_{(\mu)}}a_0\simeq 0,\ \ \partial_{r_{(\mu)}}\tilde{a}_0\simeq 0,
\ee
\JCS{where the approximate sign indicates that we have only used an approximate RG evolution to leading order in $\alpha$.}

The force between the quark and antiquark as a function of the separation can be found solving for $p$ in \eqref{eq:length1} and \eqref{eq:length2}. When $L\to\infty$, the leading corrections are
\be\label{eq:fx1}
\cF_x\simeq \frac{R^2}{2\pi\alpha'}  \frac{c_0^2}{L^2}\left[1+\frac{2 a_0}{c_0}\left(\frac{c_0 R}{L}\right)^3+\frac{2a_{\Delta-d}}{c_0}\left(\frac{c_0 R}{L}\right)^{2(\Delta-d)}\right], \ \ \Delta-d\neq \frac{3}{2},
\ee
or
\be\label{eq:fx2}
\cF_x=\frac{R^2}{2\pi\alpha'} \frac{c_0^2}{L^2}\left[ 1+\frac{2\tilde{a}_0}{c_0}\left(\frac{c_0 R}{L}\right)^3+\frac{2a_{3/2}}{c_0}\left(\frac{c_0 R}{L}\right)^3 \log\left(\frac{c_0^2 R^2}{p_0 L^2} \right)\right], \ \ \Delta-d \JCS{=} \frac{3}{2},
\ee
These expressions are valid for any RG flow flowing to an IR fixed point having a domain wall geometry as holographic dual.
\JCS{In both these expressions the first term contains the expected conformal length-dependence of the force of the IR CFT, while the last term encodes the contribution from the irrelevant operator that deforms that CFT. These two contributions are solely determined from infrared physics, once the scaling dimension and coupling of the operator are known.  The second term is more interesting, since its $L$ dependence is universal. Independently of the details of the RG-flow, provided that the holographic theory flows to an IR fixed point, there is a contribution to the force between the quark-antiquark pair that behaves as $L^{-5}$. Since the value of  $a_0$, $\tilde{a}_0$ are determined from the RG flow, at long distances all the information of the UV theory is hidden in the coefficient of this universal contribution. }

%%%%%%%%%%%%%%%%%%%%%%%%%%%%%%%%%%%%%%
\subsection{Defect theory interpretation}
%%%%%%%%%%%%%%%%%%%%%%%%%%%%%%%%%%%%%%
\JCS{The universal contribution identified above is intriguing since, a priori, it would have been hard to guess from the IR theory alone. In this subsection we will clarify its origin by analyzing the possible contributions to the quark-antiquark potential when considering the Wilson line as  a defect and studying its fluctuations. }
\rr{
In order to help us identify the origin of the universal contribution, let us first consider an arbitrary $AdS$ geometry
\be
ds^2\simeq dr^2+e^{2r/R}\eta_{\mu\nu}dx^\mu dx^\nu.
\ee
Inside this geometry we introduce a string extending straight along the radial direction and ending at the asymptotic boundary. The induced metric on the string is $AdS_2$
\be
ds_2=h_{ab}d\sigma^a d\sigma^b\simeq  d\sigma^2-e^{2\sigma/R}d\tau^2.
\ee
The string action for fluctuations $\delta x(\tau,\sigma)$ around the straight profile can be expanded to quadratic order
\be
S_{NG}\simeq \frac{1}{2\pi\alpha'}\int d^2\sigma\, e^{\sigma/R}\left( -1+\frac{1}{2}\left( (\partial_\tau \delta x)^2-e^{2\sigma/R}(\partial_\sigma \delta x)^2\right)\right).
\ee 
Using the field redefinition 
\be
\delta x=R e^{-\sigma/R} \varphi,
\ee
we can rewrite this as the action for a scalar field in $AdS_2$
\be
\label{eq:Sflucsstraight}
S_{NG}\simeq -\frac{R^2}{2\pi\alpha'}\int d^2\sigma\, \sqrt{-h}\left( 1+\frac{1}{2}\left(h^{ab}\partial_a \varphi\partial_b\varphi+m^2\varphi^2\right) \right),
\ee 
where the mass is
\be
m^2R^2=2.
\ee
Following the usual AdS/CFT dictionary, the $AdS_2$ geometry on the string is dual to a $0+1$ dimensional CFT, and the field $\varphi$ is dual to an operator $\cO_\varphi$ of dimension $\Delta$ such that $m^2R^2=\Delta(\Delta-1)$. The root $\Delta=-1$ corresponds to changes in the position of the Wilson loop at the boundary $\delta x(\infty)\neq 0$, so it can be naturally identified with changes in the quark-antiquark separation along the $x$ direction $\cO_\phi=\hat{L}$. The operator corresponding 
\JCS{to $\Delta =2$}
 can be identified from the variation of the expectation of the Wilson loop with respect to changes in the trajectory. For instance, for a BPS loop in $\cN=4$ SYM,
\be\label{eq:elecfield}
\frac{\delta }{\delta x^\mu} \log\vev{\cW}\propto \frac{1}{\vev{\cW}}\vev{ {\rm Tr}\,\left[\left(F_{\mu\nu}[x]\dot{x}^\nu+ D_\mu^\perp\Phi^I[x]\theta_I|\dot{x}| \right)e^{i\oint (A+\theta\cdot \Phi)}\right]}\equiv E_\mu^\perp,
\ee
where $F_{\mu\nu}$ is the field strength of the gauge fields. Generalizing the expression above, the $\Delta=2$ operator  can be identified as the electric field strength produced in the $x$ direction $\cO_\varphi=\hat{E}_x^\perp$. Adding to the straight string a boundary action of the form \eqref{eq:boundact} would introduce a double-trace deformation $\sim \cO_\varphi^2$ of the $0+1$ dimensional CFT that would trigger a flow to a different fixed point in the IR (see e.g. \cite{Faulkner:2010jy}). As usual, the flow would be between the alternative quantization in the UV, for which the double trace deformation is relevant $\Delta_{\cO^2}=-2$, to the normal quantization in the IR, where the double-trace deformation is irrelevant $\Delta_{\cO^2}=4$.

Now let us go back to our setup. Even though the geometry deviates from $AdS$ at large values of the radial coordinate $r\gtrsim r_{(M)}$, deep in the interior of the space, $r\ll r_{(M)}$, the geometry again approaches $AdS$, since we are considering a theory with an IR fixed point. Furthermore, in the long distance limit we have considered, the separation between the endpoints is so large that the string profile is approximately straight close to the cutoff $r_{(\mu)} \ll r_{(M)}$. As a consequence, the $AdS_2/CFT_1$ map for the string is expected work in this region and the small string fluctuations close to that cut-off are governed by the same action \eqref{eq:Sflucsstraight}. Therefore, from the point of view of the IR CFT, those fluctuations posses the same scaling dimensions and the same dual operators that we have identified in the pure $AdS$ geometry.  In particular, the boundary action \eqref{eq:boundact} corresponds to a double-trace deformation $\sim \cO_\varphi^2$ and, as the string below the cutoff lays in the IR region, the dimension of the double-trace deformation must be irrelevant and, as a consequence,  $\Delta_{\cO^2}=4$. From the point of view of the IR effective description, the boundary action triggers a flow that at higher energy scales drives away the theory from the IR CFT. As a consequence, the double trace deformation will give a contribution to the potential whose dependence on $L$ will be fixed by conformal invariance
\be
\Delta V_{q\bar{q}} \propto c_{E^2}\vev{E_x^2}\sim  \frac{c_{E^2}}{L^4}.
\ee
This will give a contribution to the force $\Delta\cF_x=-\partial_L\Delta V_{q\bar{q}} \sim 1/L^5$, whose dependence on $L$ agrees with the universal term in \eqref{eq:fx1} and \eqref{eq:fx2}.
}
%%%%%%%%%%%%%%%%%%%%%%%%%%%%%%%%%%%%%%
\section{Confining theories}\label{sec:conf}
%%%%%%%%%%%%%%%%%%%%%%%%%%%%%%%%%%%%%%

\JCS{The second application of our effective description of Wilson loops in the IR  is to study the long distance quark-antiquark potential in confining theories. One of the defining
characteristics of that type of theories is that Wilson loops follow an area law.\footnote{If the theory contains degrees of freedom in the fundamental color representation this is only true in the large-$N$ limit.} This implies that the heavy quark potential grows linearly at sufficiently large distances $V_{q\bar q} = \sigma_s L$, with $\sigma_s$ the string tension. Going beyond this leading behavior, in this second application we will study the leading correction to this potential at long distance in holographic confining theories, which will be sensitive to characteristic features of the holographic description of confinement.  
 }

In the holographic dual description \JCS{of confinement} the string dual to the Wilson loop reaches a region of space where its tension remains fixed as the separation between the endpoints at the asymptotic boundary is increased. \JCS{Although there are several different realizations of confining theories in holography, \footnote{
 For the general conditions on the metric  under which the Wilson loop will show area law in a given holographic model see \cite{Kinar:1998vq}.} in this work we will focus on two model examples}.
The first example is the WQCD model \cite{Witten:1998zw}, consisting of the gravity dual of a stack of D4 branes wrapped around a compact direction with supersymmetry-breaking boundary conditions. At weak coupling the theory is expected to flow in the IR to pure Yang-Mills. At strong coupling there is really no separation of scales between the gapped four-dimensional modes and the Kaluza-Klein modes corresponding to excitations along the fifth direction. Nevertheless the model captures some of the properties of a confining theory, including an area law for the Wilson loop \cite{Brandhuber:1998er}. The second example of a holographic dual to a confining theory is the KS model \cite{Klebanov:2000hb}, which is dual to a non-conformal $\cN=1$ supersymmetric $SU(M_c)\times SU(M_c+N_c)$ gauge theory. In the IR the theory flows to  $SU(N_c)$ $\cN=1$ super Yang-Mills and also exhibits an area law for the Wilson loop \cite{Loewy:2001pq}. 

In both examples, the area law behavior of the Wilson loop can be traced to the properties of the dual geometry in the interior. Rather than having an infinite throat, as for an $AdS$ space, or a horizon, the geometry ends smoothly at a fixed position in the radial direction when a cycle in the internal space of the geometry collapses to zero size. For the WQCD model, the cycle is the circle corresponding to the fifth direction on the dual D4 branes, while for the KS model, it is a two-cycle in the internal geometry transverse to the field theory directions. For strings with a small separation between its endpoints, the string is hanging far from the point where the space ends. Increasing the separation will make the string go deeper in the geometry, until it reaches the end of space. Since it cannot penetrate further, a larger separation of the endpoints will result in having an increasing stretch of the string lying at the bottom of space. As the tension of the string is finite in this region, the action of the string increases linearly with the separation and produces the area law. \JCS{ As we will see, corrections to the area law emerge as a consequence of the piece of the dual string extended along the radial direction throughout all the geometry}.

%%%%%%%%%%%%%%%%%%%%%%%%%%%%%%%%%%%%%%
\subsection{WQCD model}
%%%%%%%%%%%%%%%%%%%%%%%%%%%%%%%%%%%%%%

In the WQCD model the string-frame metric is usually given as
\be
ds^2=\left(\frac{U}{R}\right)^{3/2}\eta_{\mu\nu}dx^\mu dx^\nu+\left(\frac{U}{R}\right)^{3/2}\left(1-\frac{U_M^3}{U^3} \right)d\varphi^2+\left(\frac{R}{U}\right)^{3/2}\frac{dU^2}{1-\frac{U_M^3}{U^3}}.
\ee
The map to field theory quantities is
\be
U_M=\frac{2}{9}\lambda_{YM} M \alpha', \ \ R^3=\frac{1}{2 M}\lambda_{YM} \alpha'
\ee
Where $\lambda_{YM}$ is the 't Hooft coupling and $M$ is the scale of KK modes along the compact direction and determines the scale of glueball masses. The change of variables $U=r^4/(2^8 R^3)$ puts the metric in ``domain wall'' form \eqref{eq:domainwall} with 
\be
e^{2A(r)}=\left(\frac{r}{4R}\right)^6, \ \ f(r)=1-\frac{{r_{(M)}}^{12}}{r^{12}},
\ee
\JCS{where $r_{(M)}=4 R^{3/4} U_M^{1/4}$ is the position at which the geometry ends. Having expressed the metric in this way, we can find the relation between the force, $p$, and the quark-antiquark separation $L$, using the relations in \eqref{eq:stringcond}}

From the first condition in \eqref{eq:stringcond}, \JCS{ the relation between the force and the minimum string position, $\sigma_*$}, is
\be
p= \left( \frac{\sigma_*}{4R}\right)^6.
\ee
\JCS{As we have already explained, confinement implies that}
for a large enough separation between the string endpoints, $\sigma_*$ is close to the end of the geometry $\sigma_*=r_{(M)}+\delta r$,
\be
\delta r \simeq \frac{p-p_M}{6 p_M}r_{(M)},\  \ p_M= \left( \frac{r_{(M)}}{4R}\right)^6=\left( \frac{U_M}{R}\right)^{3/2}.
\ee
From \eqref{eq:stringcond}, the quark-antiquark separation is
\be
L=2a_{(\mu)} p+2p \int_{\sigma_*}^{r_{(\mu)}} d\sigma\,I(\sigma),
\ee
where
\be
I(\sigma)=\frac{e^{-3A}}{\sqrt{f}\sqrt{1-e^{-4A}p^2}}.
\ee
The integral can be \rr{calculated} analytically. Expanding for $r_{(\mu)}\gg r_{(M)},\sigma_*$ and $\delta r/r_{(M)}\ll 1$ one finds
\be
 \int_{\sigma_*}^{r_{(\mu)}}d\sigma\, I(\sigma)=-\frac{R}{2}\left[\left( \frac{4R}{r_{(\mu)}}\right)^8+\frac{2}{3}\left( \frac{4R}{r_{(M)}}\right)^8\log\left( e^{\frac{\pi}{2\sqrt{3}}} \frac{\delta r}{\sqrt{3} r_{(M)}}\right) \right].
\ee
The quark-antiquark separation in terms of $p$ is
\be\label{eq:length1b}
L=R\left[ -c_0  p\log\left(\frac{p-p_M}{p_0}\right)+ a_0 p\right],
\ee
where the coefficients are
\be
\begin{split}
&c_0=\frac{2}{3}\left( \frac{4R}{r_{(M)}}\right)^8,\\
&a_0=2 a_{(\mu)}-\left( \frac{4R}{r_{(\mu)}}\right)^8,\\
&p_0= 6\sqrt{3}p_Me^{-\frac{\pi}{2\sqrt{3}}} .
\end{split}
\ee
\JCS{As in our first application, only one of these coefficients. $a_0$, depend explicitly on the cut-off. This reveals that its value is sensitive to UV physics, unlike the other two coefficients, that depend on quantities defined at the bottom of the geometry. However, taking into account the RG-flow equation $a_{(\mu)}$ \eqref{eq:rgflowam}, the coefficient $a_0$ is, as expected, independent of the cut-off. Indeed, expanding in $r_{(M)}/r_{(\mu)} \ll 1$, one can show that}
\be
\partial_{r_{(\mu)}}a_0\simeq 0.
\ee
The force between the quark and antiquark as a function of the separation can be found solving for $p$ in \eqref{eq:length1b}. When $L\to\infty$, the leading correction is
\be
p\simeq p_M+p_0 e^{-a_0/c_0} e^{-\frac{L}{c_0 p_M R}}.
\ee
Then, the force is
\be\label{eq:forceconf}
\cF_x=\sigma_s\left( 1+q_M e^{-M L}\right),
\ee
where the string tension and the coefficient of the exponential term are
\be
\sigma_s=\frac{p_M}{2\pi \alpha'}=\frac{2}{27\pi}\lambda_{YM} M^2 ,\ \ q_M=6\sqrt{3}e^{-\frac{\pi}{2\sqrt{3}}}e^{-a_0/c_0}.
\ee
The slope of the exponential decay $M$ is determined by the IR theory, but the amplitude $q_M$ depends on $a_0$, which contains information about the RG flow above the cutoff. It should be noted that exponential corrections to the leading order behavior were already found e.g.~\cite{Kinar:1998vq,Nunez:2009da}, but the connection with the scale of glueball masses was not made.
\JCS{The appearance of these corrections, that are  non-perturbative in the $1/L$ expansion, are not just a feature of this particular model. As we will see they also appear in the KS model, which we analyze next.  }

%%%%%%%%%%%%%%%%%%%%%%%%%%%%%%%%%%%%%%
\subsection{KS model}
%%%%%%%%%%%%%%%%%%%%%%%%%%%%%%%%%%%%%%
In the KS model the metric is slightly more complicated. Separating the field theory directions from the internal ones it takes the form
\be
ds^2=h^{-1/2}(\tau)\eta_{\mu\nu}dx^\mu dx^\nu+h^{1/2}(\tau)ds^2_6.
\ee
The metric along the internal directions is the deformed conifold. It can be given in terms of the following basis of one-forms for the angular directions 
\be
g^1=\frac{e^1-e^3}{\sqrt{2}},\ g^2=\frac{e^2-e^4}{\sqrt{2}}, \ g^3=\frac{e^1+e^3}{\sqrt{2}}\, g^4=\frac{e^2+e^4}{\sqrt{2}}, \ g^5= e^5,
\ee
where
\be
\begin{split}
& e^1\equiv -\sin \theta_1 d\phi_1,\  e^2\equiv d\theta_1,\ e^3\equiv \cos \psi \sin\theta_2 d\phi_2-\sin\psi d\theta_2,\\
& e^4\equiv \sin \psi \sin\theta_2 d\phi_2+\cos\psi d\theta_2,\ e^5\equiv d\psi+\cos\theta_1 d\phi_1+\cos\theta_2 d\phi_2.
\end{split}
\ee
Then, the internal metric is
\be
ds_6^2=\frac{\varepsilon^{4/3}K(\tau)}{2}\left[\frac{1}{3(K(\tau))^3}\left(d\tau^2+(g^5)^2\right)+\cosh^2\frac{\tau}{2}\left((g^3)^2+(g^4)^2\right)+\sinh ^2\frac{\tau}{2}\left((g^1)^2+(g^2)^2\right)\right]
\ee
$\varepsilon^{2/3}$ has dimensions of length and sets the scale of the conifold deformation. The warp factors are
\be
K(\tau)=\frac{(\sinh 2\tau-2\tau)^{1/3}}{2^{1/3}\sinh \tau},\ \ h(\tau)=\alpha\int_{\tau}^\infty dx\frac{x\coth  x-1}{\sinh x}K(x),
\ee
where $\alpha=2(g_s N_c \alpha')^2 \varepsilon^{-8/3} $, with $g_s$ is the string coupling constant. The asymptotic boundary is at $\tau\to \infty$, although in this case the metric deviates from $AdS$ by logarithmic corrections. When $\tau\to 0$, a two-cycle in the internal metric collapses to zero size and the space terminates. The expansion of the warp factors in that limit is
\be
K(\tau)\simeq \left(\frac{2}{3}\right)^{1/3}\left( 1-\frac{\tau^2}{10}\right),\ \ h(\tau)=\left(\frac{2}{3}\right)^{1/3}\alpha\left( \hat{h}_M-\frac{\tau^2}{6}\right),
\ee
where $\hat{h}_M\simeq 0.65$.

The map to field theory quantities is 
\be
\varepsilon^{2/3}\propto  (g_s N_c \alpha') M, \ \ g_s N_c= \lambda_{YM}.
\ee
Where $\lambda_{YM}$ is the 't Hooft coupling and $M$ sets the scale of glueball masses \cite{Klebanov:2000hb,Caceres:2000qe}. 

Identifying $r=\varepsilon^{2/3}\tau$ as the domain wall coordinate, the warp factors are
\be
e^{2A(r)}=h^{-1/2}(\varepsilon^{-2/3}r), \ \ f(r)=6h^{-1/2}(\varepsilon^{-2/3}r)(K(\varepsilon^{-2/3}r))^2.
\ee
From the first condition in \eqref{eq:stringcond}, we obtain
\be
p=h^{-1/2}(\varepsilon^{-2/3}\sigma_*).
\ee
For a large enough separation between the string endpoints, $\sigma_*$ should be close to the end of the geometry $\sigma_*=\delta r\ll \varepsilon^{2/3}$,
\be
\delta r^2 \simeq \frac{12}{\hat{h}_M}\frac{p-p_M}{ p_M}\varepsilon^{4/3},\  \ p_M= h^{-1/2}(0)=\left(\frac{2}{3}\right)^{-1/6}\alpha^{-1/2} \hat{h}_M^{-1/2}.
\ee
From \eqref{eq:stringcond}, the quark-antiquark separation is
\be
L=2a_{(\mu)} p+2p \int_{\sigma_*}^{r_{(\mu)}} d\sigma\,I(\sigma,\sigma_*),
\ee
where
\be
I(\sigma,\sigma_*)=\frac{e^{-3A}}{\sqrt{f}\sqrt{1-e^{-4A}p^2}}=\frac{1}{\sqrt{6} p}\frac{h(\varepsilon^{-2/3}\sigma)}{K(\varepsilon^{-2/3}\sigma)\sqrt{h(\varepsilon^{-2/3}\sigma_*)-h(\varepsilon^{-2/3}\sigma)}}.
\ee
For values of $\sigma$ close to $\sigma_*$,
\be
I(\sigma,\sigma_*)\simeq I_0(\sigma,\sigma_*) =\frac{p_M}{p}\frac{\alpha\hat{h}_M^{3/2} \varepsilon^{2/3}}{\sqrt{\sigma^2-\sigma_*^2}}.
\ee
 For $\sigma_*=0$ the integral has a logarithmic divergence. In order to expand for small values of $\sigma_*=\delta r$, we subtract the divergent contribution and define
 \be
 \int_{\delta r}^{r_{(\mu)}}d\sigma\, I(\sigma)=\int_{\delta r}^{r_{(\mu)}}d\sigma\, \left[I(\sigma,\delta r)-I_0(\sigma,\delta r)\right]+\frac{p_M}{p}\alpha\hat{h}_M^{3/2}\varepsilon^{2/3}\log\left( \frac{r_{(\mu)}+\sqrt{r_{(\mu)}^2-\delta r^2}}{\delta r}\right).
 \ee
Expanding for small $\delta r$ we find, to leading order
\be
 \int_{\delta r}^{r_{(\mu)}}d\sigma\, I(\sigma)\simeq \int_{0}^{r_{(\mu)}}d\sigma\, \left[I(\sigma,0)-I_0(\sigma,0)\right]_{p=p_M}+\alpha\hat{h}_M^{3/2}\varepsilon^{2/3} \left(\log  (\varepsilon^{-2/3}r_{(\mu)})-\log\frac{\varepsilon^{-2/3}\delta r}{2} \right).
 \ee
The quark-antiquark separation in terms of $p$ is
\be\label{eq:length1c}
L= \varepsilon^{2/3} \left[ -c_0 p  \log\left(\frac{p-p_M}{p_0}\right)+ a_0 p \right],  
\ee
where the coefficients are
\be
\begin{split}
&c_0=\alpha \hat{h}_M^{3/2},\\
&a_0=2 \varepsilon^{-2/3} a_{(\mu)}+2\alpha\hat{h}_M^{3/2}\log(\varepsilon^{-2/3} r_{(\mu)})\\
  &+2\sqrt{h(0)}\int_{0}^{r_{(\mu)}}d\sigma\, \left[\frac{\varepsilon^{-2/3}}{\sqrt{6}}\frac{h(\varepsilon^{-2/3}\sigma)}{K(\varepsilon^{-2/3}\sigma)\sqrt{h(0)-h(\varepsilon^{-2/3}\sigma)}}-\frac{\alpha \hat{h}_M^{3/2}}{\sigma}\right],\\
&p_0= p_M\hat{h}_M/3.
\end{split}
\ee
Only the coefficient $a_0$ has a dependence on the cutoff. Expanding the RG-flow equations for $a_{(\mu)}$ \eqref{eq:rgflowam} for $\varepsilon^{-2/3} r_{(\mu)} \gg 1$, one can show that
\be
\partial_{r_{(\mu)}}a_0\simeq 0.
\ee
The force between the quark and antiquark as a function of the separation can be found solving for $p$ in \eqref{eq:length1c}. When $L\to\infty$, the leading corrections are
\be
p\simeq p_M+p_0 e^{-a_0/c_0} e^{-\frac{\varepsilon^{-2/3}L}{c_0 p_M}}.
\ee
We define the glueball mass scale $M$ as
\be
M=\frac{\varepsilon^{-2/3}}{c_0 p_M}=\frac{1}{(12)^{1/6}\hat{h}_M}\frac{\varepsilon^{2/3}}{g_sN_c\alpha'}.
\ee
Then, the force  takes the same form as in \eqref{eq:forceconf}, where the string tension and the coefficient of the exponential term are
\be
\sigma_s=\frac{p_M}{2\pi \alpha'}= \frac{3^{1/6}}{2\pi}\hat{h}_M^{3/2}  \lambda_{YM} M^2 ,\ \ q_M=\frac{\hat{h}_M}{3} e^{-a_0/c_0}.
\ee

%%%%%%%%%%%%%%%%%%%%%%%%%%%%%%%%%%%%%%%%%%%%%%%%%%%%%%%
\subsection{Interpretation of the Wilson loop as a flux tube in the IR}
%%%%%%%%%%%%%%%%%%%%%%%%%%%%%%%%%%%%%%%%%%%%%%%%%%%%%%%

\JCS{A  popular and successful description of the physics of confined matter is formulated in terms of effective flux tubes, whose dynamics are determined by massless transverse fluctuations in the field theory directions. These objects describe the configuration of the color gauge 
fields sourced by the heavy quark and antiquark that run along the Wilson loop. 
According to this picture,}
in a confining theory, and for a large enough separation compared to the confining scale, the color flux between the two sources is concentrated in a tube that extends from one to the other. In principle, flux tubes in the absence of sources also describe dynamical excitations of the gauge theory, but they have to be either closed or ending on dynamical quarks (or other colored particles). The Wilson loop that determines the quark-antiquark potential is then determined by the properties of the flux tube in the IR theory. We will now interpret the result obtained for the force \eqref{eq:forceconf} using the holographic Wilsonian renormalisation in terms of a modification of the effective theory of the flux tube.

\JCS{This picture emerges naturally in holography. A long classical string at the bottom of the dual geometry realizes a flux tube in the field theory. 
These elongated objects posses their own excitations, which are determined by the vibrational modes of the string in the different direction of the gravity theory.}
Fluctuations in the transverse field theory directions are massless and are determined by the four-dimensional Nambu-Goto action with string tension $\sigma_s$. In addition, there are massive fluctuations corresponding to perturbations away from the bottom, which from the point of view of the field theory dual should be interpreted as internal excitations of the flux tube.  The piece of the string that connects the parallel paths of the rectangular Wilson loop and is close to the bottom of the geometry has the same properties as a flux tube.

\JCS{To avoid subtleties with boundary conditions, let us characterize those fluctuations for a flux tube wrapping a compact spacial direction $x$ of length $L$.
This sourceless flux-tube excitation is realized by a closed string at the bottom of the geometry ( $U=U_M$ in WQCD or $\tau=0$ in KS). To better characterize the fluctuation dynamics of the flux-tube, we will use a series of changes of coordinates in each of the two models we have considered that provide a simple Lagrangian for small (quadratic) fluctuation around the classical string. }

In the WQCD model, we will use the following change of coordinates \be
U=U_M\left(1+c_\rho\rho^2+c_\rho^{(2)}\rho^4\right),\ \ \varphi=c_\theta \theta.
\ee
For $\rho\to 0$, if one fixes
\be
c_\rho=\frac{3}{4R^{3/2}U_M^{1/2}},\ \ c_\rho^{(2)}=-\frac{3}{32 R^2 U_M}, \ \ c_\theta=\frac{2}{3}\frac{R^{3/2}}{U_M^{1/2}},
\ee
the metric is, to leading order,
\be
ds^2=p_M\left(1+\frac{\rho^2}{\rho_M^2}\right)\eta_{\mu\nu}dx^\mu dx^\nu+d\rho^2+\rho^2 d\theta^2,
\ee
where
\be
\rho_M^2=\frac{8}{9}R^{3/2}U_M^{1/2}.
\ee
The periodicity of $\theta$ is
\be
\theta\sim \theta+2\pi\ \Rightarrow \ \varphi\sim \varphi+\beta_\varphi,\ \ \beta_\varphi=\frac{2\pi}{c_\theta}.
\ee
We now change to Cartesian coordinates
\be
d\rho^2+\rho^2 d\theta^2=dV^2+dW^2,
\ee
such that the metric is
\be
ds^2=p_M\left(1+\frac{V^2+W^2}{\rho_M^2}\right)\eta_{\mu\nu}dx^\mu dx^\nu+dV^2+dW^2.
\ee

Similarly, in the KS model we expand the metric for small values of $\tau$ and pick the directions transverse to the three-cycle
\be
\theta_1=\theta_2=\theta,\ \ \phi_1=-\phi_2=\phi,\ \ \psi=\pi.
\ee
The metric takes the form
\be
ds^2=p_M\left( 1+\frac{\tau^2}{12\hat{h}_M}\right)\eta_{\mu\nu}dx^\mu dx^\nu+ \frac{\varepsilon^{4/3}}{2(12)^{1/3}p_M}\left( d\tau^2+\tau^2(d\theta^2+\sin^2\theta d\phi^2)\right).
\ee
We see explicitly that the two-cycle collapses smoothly to zero size at $\tau=0$. We now introduce Cartesian coordinates to describe the region around $\tau=0$
\be
d\tau^2+\tau^2(d\theta^2+\sin^2\theta d\phi^2)=2(12)^{1/3}p_M \varepsilon^{-4/3} \left(dZ^2+dV^2+dW^2\right),
\ee
where we have rescaled the $\tau$ coordinate in such a way that the metric takes the form
\be
ds^2=p_M\left(1+\frac{Z^2+V^2+W^2}{\rho_M^2}\right)\eta_{\mu\nu}dx^\mu dx^\nu+dZ^2+dV^2+dW^2.
\ee
Where, in this case
\be
\rho_M^2=(12)^{2/3}\hat{h}_M\frac{\varepsilon^{4/3}}{2p_M} .
\ee

Denoting the field theory directions as $x^\mu=(t,x,y,z)$, we allow the string to fluctuate in the transverse directions $y$ and $W$, keeping $z=0$ and $V=0$ (and $Z=0$ for KS) fixed. In the static gauge, the embedding functions of the string are
\be
X^0=\tau,\ \ X^1=\sigma,\ \ X^2=y(\tau,\sigma),\ \ X^3=0,\ \ (Z=0),\ \ V=0,\ \ W=W(\tau,\sigma).
\ee
The induced metric is
\be
\begin{split}
&g_{\tau\tau}=-\left[p_M\left(1+\frac{W^2}{\rho_M^2}\right)\left(1-\dot{y}^2\right)-\dot{W}^2\right],\\
&g_{\sigma\sigma}=\left[p_M\left(1+\frac{W^2}{\rho_M^2}\right)\left(1+(y')^2\right)+(W')^2\right],\\
&g_{\tau\sigma}=p_M\left(1+\frac{W^2}{\rho_M^2}\right)\dot{y}y'+\dot{W}W',
\end{split}
\ee
where $\dot{y}=\partial_\tau y$, $y'=\partial_\sigma y$, etc. Expanding to second order in the fluctuations, the determinant is
\be
\sqrt{-g}=p_M\left(1-\frac{1}{2}\dot{y}^2+\frac{1}{2}(y')^2+\frac{1}{\rho_M^2} W^2\right)-\frac{1}{2}\dot{W}^2+\frac{1}{2}(W')^2.
\ee
Changing the normalization
\be
W=\sqrt{p_M}\chi,
\ee
we find
\be
\sqrt{-g}=p_M\left(1-\frac{1}{2}\dot{y}^2+\frac{1}{2}(y')^2-\frac{1}{2}\dot{\chi}^2+\frac{1}{2}(\chi')^2+\frac{m^2}{2} \chi^2\right).
\ee
Where
\be
m^2=\frac{2p_M}{\rho_M^2}, \ \ m^2_{WQCD}=\frac{9}{4}\frac{U_M}{R^3},\ \ m^2_{KS}=\frac{1}{(12)^{1/3} \hat{h}_M^2}\frac{\varepsilon^{4/3}}{(g_s N_c\alpha')^2}.
\ee
In both cases the mass of the string mode is of the same order as the mass of the glueballs $m^2=M^2$. The action for the fluctuations to quadratic order is
\be
S_{NG}=-\frac{1}{2\pi\alpha'}\int d^2 \sigma\sqrt{-g}\simeq \sigma_s\int d^2\sigma\left(-1+\frac{1}{2}\dot{y}^2-\frac{1}{2}(y')^2+\frac{1}{2}\dot{\chi}^2-\frac{1}{2}(\chi')^2-\frac{M^2}{2} \chi^2 \right).
\ee
Therefore, $M$ is the mass of the internal excitation of a flux tube in the IR theory. For the WQCD model, the same quadratic action was derived in \cite{Bigazzi:2004ze}, where one can also find the quadratic action for world-sheet fermions.
\JCS{The fluctuations of the flux tube between the quark-antiquark pair are also governed by the same quadratic Lagrangian.}

We now consider an open flux tube of length $L$. This is described in the holographic dual by a string satisfying Neumann boundary conditions at the endpoints. In this particular case, a string extended along the $x$ direction at the bottom of the geometry $W=0$ is a solution to the classical action satisfying the boundary conditions. 
\JCS{This excitation does not describe a Wilson loop, since the endpoints of the flux do not connect to the boundary. However, for large inter-quark separation, it approximates well the dual string configuration.}
In order to describe a Wilson loop, \JCS{we must modify this string such that it bends towards the boundary. Far from the edges, where the inter-quark string profile lies close to the bottom of the geometry, this can be described by a small perturbation of the open flux tube string induced by additional static sources for the massive mode at the edges of the string, }
\be
S=S_{NG}-\frac{1}{2\pi\alpha'}\int d^2\sigma W \cJ_W,
\ee
where
\be
 \cJ_W=-\sqrt{p_M}q_\chi\left(\delta(\sigma)+\delta(\sigma-L) \right).
\ee
The equation of motion for the massive mode is, for a static configuration
\be
-\chi''+M^2\chi=q_\chi\left(\delta(\sigma)+\delta(\sigma-L) \right).
\ee
This can be solved by using a Green's function, which for Neumann boundary conditions is
\be
G(\sigma,\sigma')= \frac{\Theta(\sigma-\sigma')\chi_2(\sigma)\chi_1(\sigma')+\Theta(\sigma'-\sigma)\chi_1(\sigma)\chi_2(\sigma')}{M\sinh(ML)}.
\ee
Where
\be
\chi_1=\cosh(M \sigma), \ \ \chi_2=\cosh(M(\sigma-L))
\ee
The solution is
\be
\chi=q_\chi\int_0^L d\sigma' G(\sigma,\sigma')\left(\delta(\sigma')+\delta(\sigma'-L) \right)=q_\chi\frac{\cosh(M\sigma)+\cosh(M(L-\sigma))}{M \sinh(M L)}.
\ee

\JCS{Far from the string edges, the quark-antiquark string is well described by the classical open flux tube configuration plus the massive mode perturbation $\chi$, since outside the endpoints the solution satisfies the classical string equations. We can then use the massive mode profile to determine the radial position of the lowest point of the string, in this case at $\sigma=L/2$, which, in turn, via \eqref{eq:pval} determines the force between the quark and antiquark to be }. 
\be
p=p_M\left( 1+2 q_\chi^2 e^{-ML}\right).
\ee
This reproduces \eqref{eq:forceconf} provided the strength of the source is tuned to the right value
\be
q_\chi=\sqrt{\frac{q_M}{2}}.
\ee

\JCS{The analysis we have just performed shows that the exponential corrections we have found in the quark-aniquark potential are due to non-vanishing excitations of an internal massive mode of the flux tube that connects those two sources. In holography, this mode corresponds to the fluctuations of the string along the holographic direction. For all holographic models in which confinement is associated with the closing of some cycle, as the two models we have considered, this excitation is massive and its mass is related to the glueball mass scale.  Therefore, this type of exponential correction in the heavy quark potential is a generic expectation of the holographic description of confinement.}

%%%%%%%%%%%%%%%%%%%%%%%%%%%%%%%%%%%%%%%%%%%%%%%%%%%%%%%
\section{Discussion}\label{sec:discuss}
%%%%%%%%%%%%%%%%%%%%%%%%%%%%%%%%%%%%%%%%%%%%%%%%%%%%%%%

\JCS{In this paper we have given the first steps towards  a Wilsonian RG flow analysis of Wilson loops in strongly coupled field theories which enjoy a holographic dual. We have employed this method to analyze the expectation value of the rectangular Wilson loop in those theories, which encodes the heavy  quark-antiquark potential.  In the long distance limit, the effective theory approach we have developed has allowed us to determine the heavy quark potential from IR physics, without detailed information of the UV properties of the corresponding gauge theory. The leading correction to the long distance potential in a $1/L$-expansion  is introduced via a double trace deformation of the effective one-dimensional theory localized on the world-line of the loop. 
The coefficient of that deformation depends explicitly on the effective theory cut-off and satisfies a renormalisation group equation which ensures the independence of observables on the renormalisation scale. All the information on UV physics is reduced to the fixed values of the coefficients at some given scale. The procedure we have developed is systematically improvable, and
higher order corrections can be in principle introduced by adding further multitrace terms to the IR effective action, following the standard Wilsonian procedure. By applying this method to two concrete holographic examples, we have been able to determine analytically the leading corrections to the heavy quark potential, which provides us with new understanding of the long distance dynamics in those set-ups. }

\JCS{In our first example we have analyzed a strongly coupled theory that flows to a (conformal) fixed point in the IR. As dictated by conformal symmetry in the IR, the long distance heavy quark potential behaves as $1/L$, with a coefficient determined by the IR physics, which is easy to determine. Employing our Wilsonian approach, we have found that  the double trace term introduces a universal correction, independent of how the IR CFT is deformed, } that contributes to the potential as $\sim 1/L^4$ or, equivalently, as $\sim 1/L^5$ in the force \eqref{eq:fx1},\eqref{eq:fx2}. This is consistent with the expected flow in a one-dimensional defect theory between a double trace term of scaling dimension $\Delta=-2$ in the UV to a scaling dimension $\Delta=4$ in the IR. In the UV the associated single trace operator is a variation in the trajectory of the Wilson loop, while in the IR it is the transverse electric field as defined in \eqref{eq:elecfield}.

\JCS{The second example that we have studied is a set of confining holographic gauge theories. For these theories, the analytic access to the long distance properties of the potential has allowed us to better understand the properties of the potential in terms of an effective theory of flux tubes. This description appears naturally in holography, where large Wilson loops are dual to long strings that stretch along a large distance at the bottom of the geometry and lead to the characteristic linear confining potential. Holographic models predict that the effective action of a flux tube in a confining theory is given by the four-dimensional Nambu-Goto action plus a set of additional  internal massive modes, which correspond to fluctuations along the holographic direction. Using our Wilsonian approach, we have identified the contribution of those modes to the heavy quark potential, that appear as an exponentially decaying therm $\sim e^{-ML}$, where $M$ is of the same order as the glueball masses. The information about the UV physics in this case is hidden in the factor multiplying the exponential correction.}

\JCS{The exponential term can be understood as originating from the profile of a massive mode induced by sources localized at the endpoints. The profile of the massive mode maps to the profile of the string along the holographic radial coordinate. This is a classical contribution (from the point of view of the effective action) to the quark-antiquark potential.\footnote{
Note that there can also be quantum corrections to the potential from integrating both massive and massless modes, the Luscher term being the most significant, see for instance \cite{Kinar:1999xu,Bigazzi:2004ze,Aharony:2009gg,Aharony:2010cx}. At very strong coupling they are suppressed by inverse powers of the 't Hooft coupling, this is the reason our classical string calculation does not capture them. At finite coupling the classical term is not necessarily dominant since it is exponentially suppressed at long distances}}
Note that while the exponential correction looks somewhat similar to the contributions from a rigidity term \cite{Polyakov:1986cs,Kleinert:1986bk,Caselle:2014eka} (see \cite{Brandt:2017yzw,Brandt:2018fft} for recent lattice calculations in $2+1$ dimensions),  those would be suppressed at strong 't Hooft coupling, and the coefficient of the exponential would decrease at  long separations as $\sim 1/L^{1/2}$, unlike the contribution we have identified. Therefore, this new contribution cannot be explained in terms of an effective action of the flux tube with massless transverse modes alone, it involves massive modes that have not been yet observed but should be present in any confining gauge/gravity dual qualitatively similar to the known examples. This opens the door to test if our understanding of confinement in gauge/gravity duality is qualitatively correct.

 \JCS{Consistently with this understanding}, it has been now established by lattice calculations that the Nambu-Goto action is a very good effective description of a flux tube in $2+1$ \cite{Bringoltz:2006gp,Brandt:2009tc,Brandt:2010bw} and $3+1$ dimensions \cite{Athenodorou:2010cs} (see \cite{Brandt:2016xsp} for reviews). 
\JCS{At long distances}, this observation may be partially expected from just effective theory arguments applied to a derivative expansion of transverse fluctuations of the flux tube, \JCS{without invoking holography}, since
 the energy of a long flux tube and its excitations have an expansion in odd powers of the length $L$, and corrections that deviate from four-dimensional Nambu-Goto can only start at $O(1/L^7)$ (or $O(1/L^5)$ for open strings) \cite{Aharony:2009gg,Aharony:2010cx,Aharony:2010db,Aharony:2011ga,Aharony:2013ipa}. 
 \JCS{
 However, lattice calculations of the spectrum of fluctuations of flux-tubes agree with that of a four-dimensional Nambu-Goto string  even at distances of order of the string length $\sigma_s L^2\sim 1$, except in some parity odd channels, where an additional mode has been observed \cite{Athenodorou:2010cs}. While the success of the four-dimensional Nambu-Goto description even in this regime is still an open problem,\footnote{
 The success of the four-dimensional Nambu-Goto description even in this regime has been attributed to the approximate integrability of the effective action \cite{Dubovsky:2012sh,Dubovsky:2013gi,Dubovsky:2014fma,Dubovsky:2015zey}, explaining also the deviations in parity odd channels from the appearance of an internal axion-like mode. It has been conjectured that in the large-$N$ limit the theory could become exactly integrable \cite{Dubovsky:2014fma}, although lattice calculations seem to disfavor this possibility \cite{Athenodorou:2017cmw}.
 }
 the existence of new modes beyond those of the four-dimensional Nambu-Goto string is expected from holography. However, the observed new mode does not correspond to the massive mode responsible for corrections to the potential, since this latter mode is parity even. It would be interesting to test whether other massive bosonic and fermionic modes of the holographic string are consistent with the additional excitations observed in the lattice. }
  
\JCS{The new mode we have identified may be directly observed by a detailed comparison of the quark-antiquark potential and the energy of a flux tube}. In confining holographic models, 
a flux tube of length $L$ has an energy which is the same as the quark-antiquark potential except without the classical contributions induced by sources at the endpoints.
Comparing the two one may be able to isolate the exponential correction, and get rid of other possible contributions such as finite size effects, which are also expected to be suppressed in the large-$N$ limit  \cite{Athenodorou:2010cs}. \JCS{Since this new massive mode is a generic expectation from holography, this analysis could provide an interesting check of gauge/gravity duality for confining theories}. 
\JCS{To further test holographic expectations, it would be interesting to develop further the effective action of a flux tube including internal massive modes, as well as to study the effect on meson Regge trajectories if a background profile of those modes is turned on. A comparison could be made with other holography-inspired models, such as a flux tubes with massive endpoints \cite{Kruczenski:2004me,Sonnenschein:2014jwa}.  We would also like  to mention that finite quark effects introduce corrections to the relation between the potential and the expectation value of the Wilson loop that depend on the chromoelectric field squared. In the context of the effective flux tube picture, these corrections where analyzed in \cite{Brambilla:2014eaa}. Nevertheless,  these have a different nature than the ones we have discussed, since the latter remain in the strictly infinite quark limit. It would be interesting to study the interplay of these two sets of corrections in the analysis of heavy meson properties. We leave these questions for future work.}

\JCS{Looking ahead, we would like to stress that our current  analysis is restricted to static heavy quark sources. It would be very interesting to develop a similar Wilsonian RG approach for Wilson loops along arbitrary curves, which would allow us to address a whole new suit of physical processes, such as the acceleration and radiation of heavy quarks. It would also be interesting to include finite temperature and density effects in the holographic description. In this way, we could address the energy loss of partons in strongly coupled plasma focussing in the IR properties of the plasma. This analysis could provide new theory input to the already existing holographically inspired analyses of jet quenching data in heavy ion collisions \cite{Ficnar:2013qxa,Chesler:2014jva,Casalderrey-Solana:2014bpa,Horowitz:2015dta,Rajagopal:2016uip}. 
In connection with this and other phenomenological applications, and as 
final remark,  we would like to mention that the Wilsonian approach we have pursued could be exploited in a semi-holographic approach,
where only IR physics are described using the gauge/gravity dual and the coefficients of the double trace deformations sensitive to UV physics is used to match with a given microscopic theory.}

%%%%%%%%%%%%%%%%%%%%%%%%%%%%%%%%%%%%%%%%%%%%%%%%%%%%%%%
\section*{Acknowledgments}
%%%%%%%%%%%%%%%%%%%%%%%%%%%%%%%%%%%%%%%%%%%%%%%%%%%%%%%

JCS would like to thank the hospitality of the HEP-group of University of Oviedo for their hospitality during the completion of this work. We would like to thank the anonymous referee for urging us to consider this problem within the general framework Hamilton-Jacobi flow of principal functions for Wilsonian renormalisation.
CH is partially supported by the Spanish grant MINECO-16-FPA2015-63667-P, the Ramon y Cajal fellowship RYC-2012-10370 and GRUPIN 18-174 research grant from Principado de Asturias. JCS is supported by  grants SGR-2017-754, FPA2016-76005-C2-1-P and MDM-2014-0367. 

\appendix

\section{RG flow evolution of the boundary action}\label{sec:Sflow}

We will follow the procedure introduced in \cite{Faulkner:2010jy}, where the evolution of the coefficients in the boundary action is derived from a Hamilton-Jacobi type of equation. We will not consider the most general type of string profiles, but we will restrict to work in the static gauge $X^r=\sigma$, $X^0=\tau$ and consider only static profiles $X^i=x^i(\sigma)$. 

The induced string metric for these configurations is
\be
ds_2^2=-e^{2A}d\tau^2+\frac{1}{f}\left(1+f e^{2A}(\partial_\sigma \vec{x})^2 \right)d\sigma^2.
\ee
Then, the determinant of the induced metric has the form 
\be
\sqrt{-h}=\frac{e^A}{\sqrt{f}}\sqrt{ 1+fe^{2A} (\partial_\sigma \vec{x})^2 }.
\ee
We introduce the conveniently normalized conjugate momenta that determine the force acting on the quarks in the dual theory
\be
p_i=-\frac{\delta \sqrt{-h}}{\delta \partial_\sigma x^i}=-\frac{e^{4A}}{\sqrt{-h}}\partial_\sigma x_i.
\ee
Therefore, the derivative of the profile is
\be\label{eq:derxRG}
\partial_\sigma x^i=-p^i e^{-4A} \sqrt{-h}=- p^i \frac{e^{-3A}}{\sqrt{f}\sqrt{1-e^{-4A}\vec{p}^2}},
\ee
and the Nambu-Goto action is going to be proportional to
\be\label{eq:dethRG}
\sqrt{-h}=\frac{e^A}{\sqrt{f}\sqrt{1-e^{-4A}\vec{p}^2}}.
\ee

We write string action as the Nambu-Goto action plus a boundary action that depends on the position of the cutoff in the radial direction $r_{(\mu)}$ and the position of the string profile at the cutoff $\vec{x}_{(\mu)}$, with $\vec{x}=0$ being the position of the quark in the dual field theory:
\be
S_{\rm string}=-\frac{1}{2\pi\alpha'}\int_{\sigma<r_{(\mu)}}  d^2\sigma\sqrt{-h}-\frac{1}{2\pi\alpha'} \int d\tau\, L_B[\vec{x}_{(\mu)},r_{(\mu)}].
\ee
Let us write down the condition that the action is stationary under changes of the position of the string at the cutoff 
\be
\delta S_{\rm string}=\frac{1}{2\pi\alpha'}\int_{\sigma<r_{(\mu)}}  d^2\sigma\, p_i \partial_\sigma \delta x^i-\frac{1}{2\pi\alpha'} \int d\tau\, \frac{\delta L_B}{\delta x_{(\mu)}^i} \delta x_{(\mu)}^i=0.
\ee
Since $\partial_\sigma p_i=0$, we get the condition
\be\label{eq:cond1}
\frac{\delta L_B}{\delta x_{(\mu)}^i} =p_i.
\ee
On the other hand, the string action and the solution for the string profile should be independent of the position of the cutoff, which gives a second condition
\be
\frac{d}{d r_{(\mu)}} S_{\rm string}=-\frac{1}{2\pi\alpha'}\int d\tau \sqrt{-h}\Big|_{\sigma=r_{(\mu)}} -\frac{1}{2\pi\alpha'} \int d\tau\left[\partial_{r_{(\mu)}}L_B+\frac{\delta L_B}{\delta x_{(\mu)}^i} \partial_{r_{(\mu)}} x_{(\mu)}^i \right]=0.
\ee
Solving for the radial derivative of the boundary action, we find the following RG flow evolution equation
\be
\partial_{r_{(\mu)}}L_B=- \sqrt{-h}\Big|_{\sigma=r_{(\mu)}}-\frac{\delta L_B}{\delta  x_{(\mu)}^i} \partial_{r_{(\mu)}} x_{(\mu)}^i .
\ee
This can be cast in the form of a Hamilton-Jacobi type of equation for the boundary action using \eqref{eq:derxRG},\eqref{eq:dethRG} and \eqref{eq:cond1}, 
\be\label{eq:HJ}
\partial_{r_{(\mu)}}L_B=-\frac{e^{A_{(\mu)}}}{\sqrt{f_{(\mu)}}}\sqrt{1-e^{-4A_{(\mu)}}\delta_{ij}\frac{\delta L_B}{\delta  x_{(\mu)}^i}\frac{\delta L_B}{\delta  x_{(\mu)}^j}  }. 
\ee
Now, to solve this equation, we try an ansatz where the boundary action admits a power expansion for small values of $x_{(\mu)}^i$
\be
L_B=M_{(\mu)}+\frac{1}{2 a_{(\mu)}}(\vec{x}_{(\mu)})^2+\cdots
\ee
Plugging this in \eqref{eq:HJ}, expanding for small $x_{(\mu)}^i$ and equating terms with equal powers of $x_{(\mu)}$, this gives the RG flow equations for the coefficients
\be\label{eq:RGa}
\partial_{r_{(\mu)}} M_{(\mu)}=-\frac{e^{A_{(\mu)}}}{\sqrt{f_{(\mu)}}}, \ \ \partial_{r_{(\mu)}}  a_{(\mu)}= -\frac{e^{-3 A_{(\mu)}}}{\sqrt{f_{(\mu)}}}.
\ee
These formulas coincide with the ones we derived before by direct inspection of the action integrated above the cutoff \eqref{eq:rgflowam}.

The advantage of this formalism is that it allows a systematic computation of the next order corrections by adding multitrace terms to the boundary action. For instance, the first subleading correction would correspond to a  quartic term 
\be
L_B=M_{(\mu)}+\frac{1}{2 a_{(\mu)}}(\vec{x}_{(\mu)})^2+\frac{1}{4 b_{(\mu)}}( (\vec{x}_{(\mu)})^2)^2+\cdots.
\ee
The condition \eqref{eq:cond1} gives the relation between the conjugate momenta and the positions at the cutoff, that depends on the coefficients of the multitrace terms
\be\label{eq:cond2}
p^i\simeq \frac{x_{(\mu)}^i}{a_{(\mu)}}+\frac{ (\vec{x}_{(\mu)})^2 x_{(\mu)}^i}{b_{(\mu)}}.
\ee
The RG flow equation for the coefficient of the quartic term can be derived by expanding to quartic order in $x_{(\mu)}^i$ the equation \eqref{eq:HJ} and collecting terms with the same factors of the string profile. A straightforward calculation shows that
\be\label{eq:RGb}
\partial_{r_{(\mu)}} b_{(\mu)}= -\frac{1}{2}\frac{e^{-7 A_{(\mu)}}}{\sqrt{f_{(\mu)}}} \frac{b_{(\mu)}^2}{a_{(\mu)}^4}-4\frac{e^{-3 A_{(\mu)}}}{\sqrt{f_{(\mu)}}} \frac{b_{(\mu)}}{a_{(\mu)}}.
\ee
For the particular case we are studying, where the string is extended along one direction, \eqref{eq:cond2} becomes
\be
p\simeq \frac{x_{(\mu)}}{a_{(\mu)}}+\frac{ x_{(\mu)}^3}{b_{(\mu)}}.
\ee
The, solving for $x_{(\mu)}$ to $O(p^3)$,
\be
x_{(\mu)}\simeq a_{(\mu)}p-\frac{a_{(\mu)}^4}{b_{(\mu)}}p^3.
\ee
Therefore, the total length $L$, as defined in \eqref{eq:stringcond}, is, to $O(p^3)$
\be
\frac{L}{2}\simeq a_{(\mu)}p-\frac{a_{(\mu)}^4}{b_{(\mu)}}p^3+p\int_{\sigma_*}^{r_{(\mu)}} \frac{e^{-3A}}{\sqrt{f}\sqrt{1-e^{-4A}p^2}}.
\ee
We can now show that $L$ is RG invariant at this order, using first \eqref{eq:RGa} and then \eqref{eq:RGb}
\be
\begin{split}
&\partial_{r_{(\mu)}} L=\\
&2p\left(\partial_{r_{(\mu)}}a_{(\mu)}-4p^2  \frac{\partial_{r_{(\mu)}}a_{(\mu)} a_{(\mu)}^3}{b_{(\mu)}} +p^2\partial_{r_{(\mu)}}b_{(\mu)}\frac{a_{(\mu)}^4}{b_{(\mu)}^2}+\frac{e^{-3A_{(\mu)}}}{\sqrt{f_{(\mu)}}}\left( 1+\frac{1}{2}e^{-4A_{(\mu)}}p^2+O\left(p^4e^{-8A_{(\mu)}}\right)\right) \right)\\
&=2p^3 \left( 4\frac{e^{-3A_{(\mu)}}}{\sqrt{f_{(\mu)}}} \frac{a_{(\mu)}^3}{b_{(\mu)}} +\partial_{r_{(\mu)}}b_{(\mu)}\frac{a_{(\mu)}^4}{b_{(\mu)}^2}+\frac{1}{2}\frac{e^{-7A_{(\mu)}}}{\sqrt{f_{(\mu)}}}+O\left(p^2e^{-11A_{(\mu)}}\right)\right)=O\left(p^5 e^{-11A_{(\mu)}}\right).
\end{split}
\ee
This procedure can in principle be extended to an arbitrary order of the multitrace corrections, until the desired precision is achieved.

\bibliographystyle{JHEP}

\bibliography{biblio}

\providecommand{\href}[2]{#2}\begingroup\raggedright\begin{thebibliography}{10}

\bibitem{Maldacena:1997re}
J.~M. Maldacena, \emph{{The large N limit of superconformal field theories and
  supergravity}}, \href{https://doi.org/10.1023/A:1026654312961}{\emph{Adv.
  Theor. Math. Phys.} {\bfseries 2} (1998) 231}
  [\href{https://arxiv.org/abs/9711200}{{\ttfamily 9711200}}].

\bibitem{Gubser:1998bc}
S.~S. Gubser, I.~R. Klebanov and A.~M. Polyakov, \emph{{Gauge theory
  correlators from noncritical string theory}},
  \href{https://doi.org/10.1016/S0370-2693(98)00377-3}{\emph{Phys.Lett.}
  {\bfseries B428} (1998) 105}
  [\href{https://arxiv.org/abs/hep-th/9802109}{{\ttfamily hep-th/9802109}}].

\bibitem{Witten:1998qj}
E.~Witten, \emph{{Anti-de Sitter space and holography}},
  {\emph{Adv.Theor.Math.Phys.} {\bfseries 2} (1998) 253}
  [\href{https://arxiv.org/abs/hep-th/9802150}{{\ttfamily hep-th/9802150}}].

\bibitem{Adams:2012th}
A.~Adams, L.~D. Carr, T.~Sch�fer, P.~Steinberg and J.~E. Thomas,
  \emph{{Strongly Correlated Quantum Fluids: Ultracold Quantum Gases, Quantum
  Chromodynamic Plasmas, and Holographic Duality}},
  \href{https://doi.org/10.1088/1367-2630/14/11/115009}{\emph{New J. Phys.}
  {\bfseries 14} (2012) 115009}
  [\href{https://arxiv.org/abs/1205.5180}{{\ttfamily 1205.5180}}].

\bibitem{DeWolfe:2013cua}
O.~DeWolfe, S.~S. Gubser, C.~Rosen and D.~Teaney, \emph{{Heavy ions and string
  theory}}, \href{https://doi.org/10.1016/j.ppnp.2013.11.001}{\emph{Prog. Part.
  Nucl. Phys.} {\bfseries 75} (2014) 86}
  [\href{https://arxiv.org/abs/1304.7794}{{\ttfamily 1304.7794}}].

\bibitem{Brambilla:2014jmp}
N.~Brambilla et~al., \emph{{QCD and Strongly Coupled Gauge Theories: Challenges
  and Perspectives}},
  \href{https://doi.org/10.1140/epjc/s10052-014-2981-5}{\emph{Eur. Phys. J.}
  {\bfseries C74} (2014) 2981}
  [\href{https://arxiv.org/abs/1404.3723}{{\ttfamily 1404.3723}}].

\bibitem{Ammon:2015wua}
M.~Ammon and J.~Erdmenger, \emph{{Gauge/gravity duality}}. Cambridge University
  Press, Cambridge, 2015.

\bibitem{2017stmc.book.....N}
H.~{Nastase}, \emph{{String Theory Methods for Condensed Matter Physics}}.
  Sept., 2017.

\bibitem{CasalderreySolana:2011us}
J.~Casalderrey-Solana, H.~Liu, D.~Mateos, K.~Rajagopal and U.~A. Wiedemann,
  \emph{{Gauge/String Duality, Hot QCD and Heavy Ion Collisions}},
  \href{https://arxiv.org/abs/1101.0618}{{\ttfamily 1101.0618}}.

\bibitem{Bhattacharyya:2008jc}
S.~Bhattacharyya, V.~E. Hubeny, S.~Minwalla and M.~Rangamani, \emph{{Nonlinear
  Fluid Dynamics from Gravity}},
  \href{https://doi.org/10.1088/1126-6708/2008/02/045}{\emph{JHEP} {\bfseries
  02} (2008) 045} [\href{https://arxiv.org/abs/0712.2456}{{\ttfamily
  0712.2456}}].

\bibitem{Iqbal:2008by}
N.~Iqbal and H.~Liu, \emph{{Universality of the hydrodynamic limit in AdS/CFT
  and the membrane paradigm}},
  \href{https://doi.org/10.1103/PhysRevD.79.025023}{\emph{Phys. Rev.}
  {\bfseries D79} (2009) 025023}
  [\href{https://arxiv.org/abs/0809.3808}{{\ttfamily 0809.3808}}].

\bibitem{Eling:2009sj}
C.~Eling and Y.~Oz, \emph{{Relativistic CFT Hydrodynamics from the Membrane
  Paradigm}}, \href{https://doi.org/10.1007/JHEP02(2010)069}{\emph{JHEP}
  {\bfseries 02} (2010) 069} [\href{https://arxiv.org/abs/0906.4999}{{\ttfamily
  0906.4999}}].

\bibitem{Faulkner:2009wj}
T.~Faulkner, H.~Liu, J.~McGreevy and D.~Vegh, \emph{{Emergent quantum
  criticality, Fermi surfaces, and AdS(2)}},
  \href{https://doi.org/10.1103/PhysRevD.83.125002}{\emph{Phys. Rev.}
  {\bfseries D83} (2011) 125002}
  [\href{https://arxiv.org/abs/0907.2694}{{\ttfamily 0907.2694}}].

\bibitem{Bredberg:2010ky}
I.~Bredberg, C.~Keeler, V.~Lysov and A.~Strominger, \emph{{Wilsonian Approach
  to Fluid/Gravity Duality}},
  \href{https://doi.org/10.1007/JHEP03(2011)141}{\emph{JHEP} {\bfseries 03}
  (2011) 141} [\href{https://arxiv.org/abs/1006.1902}{{\ttfamily 1006.1902}}].

\bibitem{Nickel:2010pr}
D.~Nickel and D.~T. Son, \emph{{Deconstructing holographic liquids}},
  \href{https://doi.org/10.1088/1367-2630/13/7/075010}{\emph{New J. Phys.}
  {\bfseries 13} (2011) 075010}
  [\href{https://arxiv.org/abs/1009.3094}{{\ttfamily 1009.3094}}].

\bibitem{Heemskerk:2010hk}
I.~Heemskerk and J.~Polchinski, \emph{{Holographic and Wilsonian
  Renormalization Groups}},
  \href{https://doi.org/10.1007/JHEP06(2011)031}{\emph{JHEP} {\bfseries 06}
  (2011) 031} [\href{https://arxiv.org/abs/1010.1264}{{\ttfamily 1010.1264}}].

\bibitem{Faulkner:2010jy}
T.~Faulkner, H.~Liu and M.~Rangamani, \emph{{Integrating out geometry:
  Holographic Wilsonian RG and the membrane paradigm}},
  \href{https://doi.org/10.1007/JHEP08(2011)051}{\emph{JHEP} {\bfseries 08}
  (2011) 051} [\href{https://arxiv.org/abs/1010.4036}{{\ttfamily 1010.4036}}].

\bibitem{Charmousis:2010zz}
C.~Charmousis, B.~Gouteraux, B.~S. Kim, E.~Kiritsis and R.~Meyer,
  \emph{{Effective Holographic Theories for low-temperature condensed matter
  systems}}, \href{https://doi.org/10.1007/JHEP11(2010)151}{\emph{JHEP}
  {\bfseries 11} (2010) 151} [\href{https://arxiv.org/abs/1005.4690}{{\ttfamily
  1005.4690}}].

\bibitem{Donos:2015gia}
A.~Donos and J.~P. Gauntlett, \emph{{Navier-Stokes Equations on Black Hole
  Horizons and DC Thermoelectric Conductivity}},
  \href{https://doi.org/10.1103/PhysRevD.92.121901}{\emph{Phys. Rev.}
  {\bfseries D92} (2015) 121901}
  [\href{https://arxiv.org/abs/1506.01360}{{\ttfamily 1506.01360}}].

\bibitem{Maldacena:1998im}
J.~M. Maldacena, \emph{{Wilson loops in large N field theories}},
  \href{https://doi.org/10.1103/PhysRevLett.80.4859}{\emph{Phys. Rev. Lett.}
  {\bfseries 80} (1998) 4859}
  [\href{https://arxiv.org/abs/hep-th/9803002}{{\ttfamily hep-th/9803002}}].

\bibitem{Rey:1998ik}
S.-J. Rey and J.-T. Yee, \emph{{Macroscopic strings as heavy quarks in large N
  gauge theory and anti-de Sitter supergravity}},
  \href{https://doi.org/10.1007/s100520100799}{\emph{Eur. Phys. J.} {\bfseries
  C22} (2001) 379} [\href{https://arxiv.org/abs/hep-th/9803001}{{\ttfamily
  hep-th/9803001}}].

\bibitem{Kiritsis:2014kua}
E.~Kiritsis, W.~Li and F.~Nitti, \emph{{Holographic RG flow and the Quantum
  Effective Action}},
  \href{https://doi.org/10.1002/prop.201400007}{\emph{Fortsch. Phys.}
  {\bfseries 62} (2014) 389} [\href{https://arxiv.org/abs/1401.0888}{{\ttfamily
  1401.0888}}].

\bibitem{Bakas:2007tm}
I.~Bakas and C.~Sourdis, \emph{{Dirichlet sigma models and mean curvature
  flow}}, \href{https://doi.org/10.1088/1126-6708/2007/06/057}{\emph{JHEP}
  {\bfseries 06} (2007) 057} [\href{https://arxiv.org/abs/0704.3985}{{\ttfamily
  0704.3985}}].

\bibitem{Witten:1998zw}
E.~Witten, \emph{{Anti-de Sitter space, thermal phase transition, and
  confinement in gauge theories}},
  \href{https://doi.org/10.4310/ATMP.1998.v2.n3.a3}{\emph{Adv. Theor. Math.
  Phys.} {\bfseries 2} (1998) 505}
  [\href{https://arxiv.org/abs/hep-th/9803131}{{\ttfamily hep-th/9803131}}].

\bibitem{Klebanov:2000hb}
I.~R. Klebanov and M.~J. Strassler, \emph{{Supergravity and a confining gauge
  theory: Duality cascades and chi SB resolution of naked singularities}},
  \href{https://doi.org/10.1088/1126-6708/2000/08/052}{\emph{JHEP} {\bfseries
  08} (2000) 052} [\href{https://arxiv.org/abs/hep-th/0007191}{{\ttfamily
  hep-th/0007191}}].

\bibitem{Girardello:1998pd}
L.~Girardello, M.~Petrini, M.~Porrati and A.~Zaffaroni, \emph{{Novel local CFT
  and exact results on perturbations of N=4 superYang Mills from AdS
  dynamics}}, \href{https://doi.org/10.1088/1126-6708/1998/12/022}{\emph{JHEP}
  {\bfseries 12} (1998) 022}
  [\href{https://arxiv.org/abs/hep-th/9810126}{{\ttfamily hep-th/9810126}}].

\bibitem{Distler:1998gb}
J.~Distler and F.~Zamora, \emph{{Nonsupersymmetric conformal field theories
  from stable anti-de Sitter spaces}},
  \href{https://doi.org/10.4310/ATMP.1998.v2.n6.a6}{\emph{Adv. Theor. Math.
  Phys.} {\bfseries 2} (1999) 1405}
  [\href{https://arxiv.org/abs/hep-th/9810206}{{\ttfamily hep-th/9810206}}].

\bibitem{Khavaev:1998fb}
A.~Khavaev, K.~Pilch and N.~P. Warner, \emph{{New vacua of gauged N=8
  supergravity in five-dimensions}},
  \href{https://doi.org/10.1016/S0370-2693(00)00795-4}{\emph{Phys. Lett.}
  {\bfseries B487} (2000) 14}
  [\href{https://arxiv.org/abs/hep-th/9812035}{{\ttfamily hep-th/9812035}}].

\bibitem{Freedman:1999gp}
D.~Z. Freedman, S.~S. Gubser, K.~Pilch and N.~P. Warner, \emph{{Renormalization
  group flows from holography supersymmetry and a c theorem}},
  \href{https://doi.org/10.4310/ATMP.1999.v3.n2.a7}{\emph{Adv. Theor. Math.
  Phys.} {\bfseries 3} (1999) 363}
  [\href{https://arxiv.org/abs/hep-th/9904017}{{\ttfamily hep-th/9904017}}].

\bibitem{Behrndt:1999ay}
K.~Behrndt, \emph{{Domain walls of D = 5 supergravity and fixpoints of N=1
  superYang Mills}},
  \href{https://doi.org/10.1016/S0550-3213(99)00773-7}{\emph{Nucl. Phys.}
  {\bfseries B573} (2000) 127}
  [\href{https://arxiv.org/abs/hep-th/9907070}{{\ttfamily hep-th/9907070}}].

\bibitem{Khavaev:2000gb}
A.~Khavaev and N.~P. Warner, \emph{{A Class of N=1 supersymmetric RG flows from
  five-dimensional N=8 supergravity}},
  \href{https://doi.org/10.1016/S0370-2693(00)01228-4}{\emph{Phys. Lett.}
  {\bfseries B495} (2000) 215}
  [\href{https://arxiv.org/abs/hep-th/0009159}{{\ttfamily hep-th/0009159}}].

\bibitem{Lu:1999bw}
H.~Lu, C.~N. Pope and T.~A. Tran, \emph{{Five-dimensional N=4, SU(2) x U(1)
  gauged supergravity from type IIB}},
  \href{https://doi.org/10.1016/S0370-2693(00)00073-3}{\emph{Phys. Lett.}
  {\bfseries B475} (2000) 261}
  [\href{https://arxiv.org/abs/hep-th/9909203}{{\ttfamily hep-th/9909203}}].

\bibitem{Cvetic:1999xp}
M.~Cvetic, M.~J. Duff, P.~Hoxha, J.~T. Liu, H.~Lu, J.~X. Lu et~al.,
  \emph{{Embedding AdS black holes in ten-dimensions and eleven-dimensions}},
  \href{https://doi.org/10.1016/S0550-3213(99)00419-8}{\emph{Nucl. Phys.}
  {\bfseries B558} (1999) 96}
  [\href{https://arxiv.org/abs/hep-th/9903214}{{\ttfamily hep-th/9903214}}].

\bibitem{Nastase:2000tu}
H.~Nastase and D.~Vaman, \emph{{On the nonlinear KK reductions on spheres of
  supergravity theories}},
  \href{https://doi.org/10.1016/S0550-3213(00)00214-5}{\emph{Nucl. Phys.}
  {\bfseries B583} (2000) 211}
  [\href{https://arxiv.org/abs/hep-th/0002028}{{\ttfamily hep-th/0002028}}].

\bibitem{Pilch:2000ue}
K.~Pilch and N.~P. Warner, \emph{{N=2 supersymmetric RG flows and the IIB
  dilaton}}, \href{https://doi.org/10.1016/S0550-3213(00)00656-8}{\emph{Nucl.
  Phys.} {\bfseries B594} (2001) 209}
  [\href{https://arxiv.org/abs/hep-th/0004063}{{\ttfamily hep-th/0004063}}].

\bibitem{Cvetic:2000nc}
M.~Cvetic, H.~Lu, C.~N. Pope, A.~Sadrzadeh and T.~A. Tran, \emph{{Consistent
  SO(6) reduction of type IIB supergravity on S**5}},
  \href{https://doi.org/10.1016/S0550-3213(00)00372-2}{\emph{Nucl. Phys.}
  {\bfseries B586} (2000) 275}
  [\href{https://arxiv.org/abs/hep-th/0003103}{{\ttfamily hep-th/0003103}}].

\bibitem{Lee:2014mla}
K.~Lee, C.~Strickland‐Constable and D.~Waldram, \emph{{Spheres, generalised
  parallelisability and consistent truncations}},
  \href{https://doi.org/10.1002/prop.201700048}{\emph{Fortsch. Phys.}
  {\bfseries 65} (2017) 1700048}
  [\href{https://arxiv.org/abs/1401.3360}{{\ttfamily 1401.3360}}].

\bibitem{Ciceri:2014wya}
F.~Ciceri, B.~de~Wit and O.~Varela, \emph{{IIB supergravity and the E$_{6(6)}$
  covariant vector-tensor hierarchy}},
  \href{https://doi.org/10.1007/JHEP04(2015)094}{\emph{JHEP} {\bfseries 04}
  (2015) 094} [\href{https://arxiv.org/abs/1412.8297}{{\ttfamily 1412.8297}}].

\bibitem{Baguet:2015sma}
A.~Baguet, O.~Hohm and H.~Samtleben, \emph{{Consistent Type IIB Reductions to
  Maximal 5D Supergravity}},
  \href{https://doi.org/10.1103/PhysRevD.92.065004}{\emph{Phys. Rev.}
  {\bfseries D92} (2015) 065004}
  [\href{https://arxiv.org/abs/1506.01385}{{\ttfamily 1506.01385}}].

\bibitem{Pilch:2000fu}
K.~Pilch and N.~P. Warner, \emph{{N=1 supersymmetric renormalization group
  flows from IIB supergravity}},
  \href{https://doi.org/10.4310/ATMP.2000.v4.n3.a5}{\emph{Adv. Theor. Math.
  Phys.} {\bfseries 4} (2002) 627}
  [\href{https://arxiv.org/abs/hep-th/0006066}{{\ttfamily hep-th/0006066}}].

\bibitem{Alday:2007he}
L.~F. Alday and J.~Maldacena, \emph{{Comments on gluon scattering amplitudes
  via AdS/CFT}},
  \href{https://doi.org/10.1088/1126-6708/2007/11/068}{\emph{JHEP} {\bfseries
  11} (2007) 068} [\href{https://arxiv.org/abs/0710.1060}{{\ttfamily
  0710.1060}}].

\bibitem{Polchinski:2011im}
J.~Polchinski and J.~Sully, \emph{{Wilson Loop Renormalization Group Flows}},
  \href{https://doi.org/10.1007/JHEP10(2011)059}{\emph{JHEP} {\bfseries 10}
  (2011) 059} [\href{https://arxiv.org/abs/1104.5077}{{\ttfamily 1104.5077}}].

\bibitem{Kinar:1998vq}
Y.~Kinar, E.~Schreiber and J.~Sonnenschein, \emph{{Q anti-Q potential from
  strings in curved space-time: Classical results}},
  \href{https://doi.org/10.1016/S0550-3213(99)00652-5}{\emph{Nucl. Phys.}
  {\bfseries B566} (2000) 103}
  [\href{https://arxiv.org/abs/hep-th/9811192}{{\ttfamily hep-th/9811192}}].

\bibitem{Skenderis:1999mm}
K.~Skenderis and P.~K. Townsend, \emph{{Gravitational stability and
  renormalization group flow}},
  \href{https://doi.org/10.1016/S0370-2693(99)01212-5}{\emph{Phys. Lett.}
  {\bfseries B468} (1999) 46}
  [\href{https://arxiv.org/abs/hep-th/9909070}{{\ttfamily hep-th/9909070}}].

\bibitem{Freedman:2003ax}
D.~Z. Freedman, C.~Nunez, M.~Schnabl and K.~Skenderis, \emph{{Fake supergravity
  and domain wall stability}},
  \href{https://doi.org/10.1103/PhysRevD.69.104027}{\emph{Phys. Rev.}
  {\bfseries D69} (2004) 104027}
  [\href{https://arxiv.org/abs/hep-th/0312055}{{\ttfamily hep-th/0312055}}].

\bibitem{Celi:2004st}
A.~Celi, A.~Ceresole, G.~Dall'Agata, A.~Van~Proeyen and M.~Zagermann, \emph{{On
  the fakeness of fake supergravity}},
  \href{https://doi.org/10.1103/PhysRevD.71.045009}{\emph{Phys. Rev.}
  {\bfseries D71} (2005) 045009}
  [\href{https://arxiv.org/abs/hep-th/0410126}{{\ttfamily hep-th/0410126}}].

\bibitem{Zagermann:2004ac}
M.~Zagermann, \emph{{N=4 fake supergravity}},
  \href{https://doi.org/10.1103/PhysRevD.71.125007}{\emph{Phys. Rev.}
  {\bfseries D71} (2005) 125007}
  [\href{https://arxiv.org/abs/hep-th/0412081}{{\ttfamily hep-th/0412081}}].

\bibitem{Skenderis:2006jq}
K.~Skenderis and P.~K. Townsend, \emph{{Hidden supersymmetry of domain walls
  and cosmologies}},
  \href{https://doi.org/10.1103/PhysRevLett.96.191301}{\emph{Phys. Rev. Lett.}
  {\bfseries 96} (2006) 191301}
  [\href{https://arxiv.org/abs/hep-th/0602260}{{\ttfamily hep-th/0602260}}].

\bibitem{deBoer:1999tgo}
J.~de~Boer, E.~P. Verlinde and H.~L. Verlinde, \emph{{On the holographic
  renormalization group}},
  \href{https://doi.org/10.1088/1126-6708/2000/08/003}{\emph{JHEP} {\bfseries
  08} (2000) 003} [\href{https://arxiv.org/abs/hep-th/9912012}{{\ttfamily
  hep-th/9912012}}].

\bibitem{Kiritsis:2016kog}
E.~Kiritsis, F.~Nitti and L.~Silva~Pimenta, \emph{{Exotic RG Flows from
  Holography}}, \href{https://doi.org/10.1002/prop.201600120}{\emph{Fortsch.
  Phys.} {\bfseries 65} (2017) 1600120}
  [\href{https://arxiv.org/abs/1611.05493}{{\ttfamily 1611.05493}}].

\bibitem{Nitti:2017cbu}
F.~Nitti, L.~Silva~Pimenta and D.~A. Steer, \emph{{On multi-field flows in
  gravity and holography}},
  \href{https://doi.org/10.1007/JHEP07(2018)022}{\emph{JHEP} {\bfseries 07}
  (2018) 022} [\href{https://arxiv.org/abs/1711.10969}{{\ttfamily
  1711.10969}}].

\bibitem{Salopek:1990jq}
D.~S. Salopek and J.~R. Bond, \emph{{Nonlinear evolution of long wavelength
  metric fluctuations in inflationary models}},
  \href{https://doi.org/10.1103/PhysRevD.42.3936}{\emph{Phys. Rev.} {\bfseries
  D42} (1990) 3936}.

\bibitem{Brandhuber:1998er}
A.~Brandhuber, N.~Itzhaki, J.~Sonnenschein and S.~Yankielowicz, \emph{{Wilson
  loops, confinement, and phase transitions in large N gauge theories from
  supergravity}},
  \href{https://doi.org/10.1088/1126-6708/1998/06/001}{\emph{JHEP} {\bfseries
  06} (1998) 001} [\href{https://arxiv.org/abs/hep-th/9803263}{{\ttfamily
  hep-th/9803263}}].

\bibitem{Loewy:2001pq}
A.~Loewy and J.~Sonnenschein, \emph{{On the holographic duals of N=1 gauge
  dynamics}}, \href{https://doi.org/10.1088/1126-6708/2001/08/007}{\emph{JHEP}
  {\bfseries 08} (2001) 007}
  [\href{https://arxiv.org/abs/hep-th/0103163}{{\ttfamily hep-th/0103163}}].

\bibitem{Nunez:2009da}
C.~Nunez, M.~Piai and A.~Rago, \emph{{Wilson Loops in string duals of Walking
  and Flavored Systems}},
  \href{https://doi.org/10.1103/PhysRevD.81.086001}{\emph{Phys. Rev.}
  {\bfseries D81} (2010) 086001}
  [\href{https://arxiv.org/abs/0909.0748}{{\ttfamily 0909.0748}}].

\bibitem{Caceres:2000qe}
E.~Caceres and R.~Hernandez, \emph{{Glueball masses for the deformed conifold
  theory}}, \href{https://doi.org/10.1016/S0370-2693(01)00278-7}{\emph{Phys.
  Lett.} {\bfseries B504} (2001) 64}
  [\href{https://arxiv.org/abs/hep-th/0011204}{{\ttfamily hep-th/0011204}}].

\bibitem{Bigazzi:2004ze}
F.~Bigazzi, A.~L. Cotrone, L.~Martucci and L.~A. Pando~Zayas, \emph{{Wilson
  loop, Regge trajectory and hadron masses in a Yang-Mills theory from
  semiclassical strings}},
  \href{https://doi.org/10.1103/PhysRevD.71.066002}{\emph{Phys. Rev.}
  {\bfseries D71} (2005) 066002}
  [\href{https://arxiv.org/abs/hep-th/0409205}{{\ttfamily hep-th/0409205}}].

\bibitem{Kinar:1999xu}
Y.~Kinar, E.~Schreiber, J.~Sonnenschein and N.~Weiss, \emph{{Quantum
  fluctuations of Wilson loops from string models}},
  \href{https://doi.org/10.1016/S0550-3213(00)00238-8}{\emph{Nucl. Phys.}
  {\bfseries B583} (2000) 76}
  [\href{https://arxiv.org/abs/hep-th/9911123}{{\ttfamily hep-th/9911123}}].

\bibitem{Aharony:2009gg}
O.~Aharony and E.~Karzbrun, \emph{{On the effective action of confining
  strings}}, \href{https://doi.org/10.1088/1126-6708/2009/06/012}{\emph{JHEP}
  {\bfseries 06} (2009) 012} [\href{https://arxiv.org/abs/0903.1927}{{\ttfamily
  0903.1927}}].

\bibitem{Aharony:2010cx}
O.~Aharony and M.~Field, \emph{{On the effective theory of long open strings}},
  \href{https://doi.org/10.1007/JHEP01(2011)065}{\emph{JHEP} {\bfseries 01}
  (2011) 065} [\href{https://arxiv.org/abs/1008.2636}{{\ttfamily 1008.2636}}].

\bibitem{Polyakov:1986cs}
A.~M. Polyakov, \emph{{Fine Structure of Strings}},
  \href{https://doi.org/10.1016/0550-3213(86)90162-8}{\emph{Nucl. Phys.}
  {\bfseries B268} (1986) 406}.

\bibitem{Kleinert:1986bk}
H.~Kleinert, \emph{{The Membrane Properties of Condensing Strings}},
  \href{https://doi.org/10.1016/0370-2693(86)91111-1}{\emph{Phys. Lett.}
  {\bfseries B174} (1986) 335}.

\bibitem{Caselle:2014eka}
M.~Caselle, M.~Panero, R.~Pellegrini and D.~Vadacchino, \emph{{A different kind
  of string}}, \href{https://doi.org/10.1007/JHEP01(2015)105}{\emph{JHEP}
  {\bfseries 01} (2015) 105} [\href{https://arxiv.org/abs/1406.5127}{{\ttfamily
  1406.5127}}].

\bibitem{Brandt:2017yzw}
B.~B. Brandt, \emph{{Spectrum of the open QCD flux tube and its effective
  string description I: 3d static potential in SU(N = 2, 3)}},
  \href{https://doi.org/10.1007/JHEP07(2017)008}{\emph{JHEP} {\bfseries 07}
  (2017) 008} [\href{https://arxiv.org/abs/1705.03828}{{\ttfamily
  1705.03828}}].

\bibitem{Brandt:2018fft}
B.~B. Brandt, \emph{{Spectrum of the open QCD flux tube and its effective
  string description}},  2018,
  \href{https://arxiv.org/abs/1811.11779}{{\ttfamily 1811.11779}}.

\bibitem{Bringoltz:2006gp}
B.~Bringoltz and M.~Teper, \emph{{String tensions of SU(N) gauge theories in
  2+1 dimensions}}, \href{https://doi.org/10.22323/1.032.0041}{\emph{PoS}
  {\bfseries LAT2006} (2006) 041}
  [\href{https://arxiv.org/abs/hep-lat/0610035}{{\ttfamily hep-lat/0610035}}].

\bibitem{Brandt:2009tc}
B.~B. Brandt and P.~Majumdar, \emph{{Spectrum of the QCD flux tube in 3d SU(2)
  lattice gauge theory}},
  \href{https://doi.org/10.1016/j.physletb.2009.11.010}{\emph{Phys. Lett.}
  {\bfseries B682} (2009) 253}
  [\href{https://arxiv.org/abs/0905.4195}{{\ttfamily 0905.4195}}].

\bibitem{Brandt:2010bw}
B.~B. Brandt, \emph{{Probing boundary-corrections to Nambu-Goto open string
  energy levels in 3d SU(2) gauge theory}},
  \href{https://doi.org/10.1007/JHEP02(2011)040}{\emph{JHEP} {\bfseries 02}
  (2011) 040} [\href{https://arxiv.org/abs/1010.3625}{{\ttfamily 1010.3625}}].

\bibitem{Athenodorou:2010cs}
A.~Athenodorou, B.~Bringoltz and M.~Teper, \emph{{Closed flux tubes and their
  string description in D=3+1 SU(N) gauge theories}},
  \href{https://doi.org/10.1007/JHEP02(2011)030}{\emph{JHEP} {\bfseries 02}
  (2011) 030} [\href{https://arxiv.org/abs/1007.4720}{{\ttfamily 1007.4720}}].

\bibitem{Brandt:2016xsp}
B.~B. Brandt and M.~Meineri, \emph{{Effective string description of confining
  flux tubes}}, \href{https://doi.org/10.1142/S0217751X16430016}{\emph{Int. J.
  Mod. Phys.} {\bfseries A31} (2016) 1643001}
  [\href{https://arxiv.org/abs/1603.06969}{{\ttfamily 1603.06969}}].

\bibitem{Aharony:2010db}
O.~Aharony and N.~Klinghoffer, \emph{{Corrections to Nambu-Goto energy levels
  from the effective string action}},
  \href{https://doi.org/10.1007/JHEP12(2010)058}{\emph{JHEP} {\bfseries 12}
  (2010) 058} [\href{https://arxiv.org/abs/1008.2648}{{\ttfamily 1008.2648}}].

\bibitem{Aharony:2011ga}
O.~Aharony, M.~Field and N.~Klinghoffer, \emph{{The effective string spectrum
  in the orthogonal gauge}},
  \href{https://doi.org/10.1007/JHEP04(2012)048}{\emph{JHEP} {\bfseries 04}
  (2012) 048} [\href{https://arxiv.org/abs/1111.5757}{{\ttfamily 1111.5757}}].

\bibitem{Aharony:2013ipa}
O.~Aharony and Z.~Komargodski, \emph{{The Effective Theory of Long Strings}},
  \href{https://doi.org/10.1007/JHEP05(2013)118}{\emph{JHEP} {\bfseries 05}
  (2013) 118} [\href{https://arxiv.org/abs/1302.6257}{{\ttfamily 1302.6257}}].

\bibitem{Dubovsky:2012sh}
S.~Dubovsky, R.~Flauger and V.~Gorbenko, \emph{{Effective String Theory
  Revisited}}, \href{https://doi.org/10.1007/JHEP09(2012)044}{\emph{JHEP}
  {\bfseries 09} (2012) 044} [\href{https://arxiv.org/abs/1203.1054}{{\ttfamily
  1203.1054}}].

\bibitem{Dubovsky:2013gi}
S.~Dubovsky, R.~Flauger and V.~Gorbenko, \emph{{Evidence from Lattice Data for
  a New Particle on the Worldsheet of the QCD Flux Tube}},
  \href{https://doi.org/10.1103/PhysRevLett.111.062006}{\emph{Phys. Rev. Lett.}
  {\bfseries 111} (2013) 062006}
  [\href{https://arxiv.org/abs/1301.2325}{{\ttfamily 1301.2325}}].

\bibitem{Dubovsky:2014fma}
S.~Dubovsky, R.~Flauger and V.~Gorbenko, \emph{{Flux Tube Spectra from
  Approximate Integrability at Low Energies}},
  \href{https://doi.org/10.1134/S1063776115030188}{\emph{J. Exp. Theor. Phys.}
  {\bfseries 120} (2015) 399}
  [\href{https://arxiv.org/abs/1404.0037}{{\ttfamily 1404.0037}}].

\bibitem{Dubovsky:2015zey}
S.~Dubovsky and V.~Gorbenko, \emph{{Towards a Theory of the QCD String}},
  \href{https://doi.org/10.1007/JHEP02(2016)022}{\emph{JHEP} {\bfseries 02}
  (2016) 022} [\href{https://arxiv.org/abs/1511.01908}{{\ttfamily
  1511.01908}}].

\bibitem{Athenodorou:2017cmw}
A.~Athenodorou and M.~Teper, \emph{{On the mass of the world-sheet 'axion' in
  $SU(N)$ gauge theories in 3$+$1 dimensions}},
  \href{https://doi.org/10.1016/j.physletb.2017.05.082}{\emph{Phys. Lett.}
  {\bfseries B771} (2017) 408}
  [\href{https://arxiv.org/abs/1702.03717}{{\ttfamily 1702.03717}}].

\bibitem{Kruczenski:2004me}
M.~Kruczenski, L.~A. Pando~Zayas, J.~Sonnenschein and D.~Vaman, \emph{{Regge
  trajectories for mesons in the holographic dual of large-N(c) QCD}},
  \href{https://doi.org/10.1088/1126-6708/2005/06/046}{\emph{JHEP} {\bfseries
  06} (2005) 046} [\href{https://arxiv.org/abs/hep-th/0410035}{{\ttfamily
  hep-th/0410035}}].

\bibitem{Sonnenschein:2014jwa}
J.~Sonnenschein and D.~Weissman, \emph{{Rotating strings confronting PDG
  mesons}}, \href{https://doi.org/10.1007/JHEP08(2014)013}{\emph{JHEP}
  {\bfseries 08} (2014) 013} [\href{https://arxiv.org/abs/1402.5603}{{\ttfamily
  1402.5603}}].

\bibitem{Brambilla:2014eaa}
N.~Brambilla, M.~Groher, H.~E. Martinez and A.~Vairo, \emph{{Effective string
  theory and the long-range relativistic corrections to the quark-antiquark
  potential}}, \href{https://doi.org/10.1103/PhysRevD.90.114032}{\emph{Phys.
  Rev.} {\bfseries D90} (2014) 114032}
  [\href{https://arxiv.org/abs/1407.7761}{{\ttfamily 1407.7761}}].

\bibitem{Ficnar:2013qxa}
A.~Ficnar, S.~S. Gubser and M.~Gyulassy, \emph{{Shooting String Holography of
  Jet Quenching at RHIC and LHC}},
  \href{https://doi.org/10.1016/j.physletb.2014.10.016}{\emph{Phys. Lett.}
  {\bfseries B738} (2014) 464}
  [\href{https://arxiv.org/abs/1311.6160}{{\ttfamily 1311.6160}}].

\bibitem{Chesler:2014jva}
P.~M. Chesler and K.~Rajagopal, \emph{{Jet quenching in strongly coupled
  plasma}}, \href{https://doi.org/10.1103/PhysRevD.90.025033}{\emph{Phys. Rev.}
  {\bfseries D90} (2014) 025033}
  [\href{https://arxiv.org/abs/1402.6756}{{\ttfamily 1402.6756}}].

\bibitem{Casalderrey-Solana:2014bpa}
J.~Casalderrey-Solana, D.~C. Gulhan, J.~G. Milhano, D.~Pablos and K.~Rajagopal,
  \emph{{A Hybrid Strong/Weak Coupling Approach to Jet Quenching}},
  \href{https://doi.org/10.1007/JHEP09(2015)175,
  10.1007/JHEP10(2014)019}{\emph{JHEP} {\bfseries 10} (2014) 019}
  [\href{https://arxiv.org/abs/1405.3864}{{\ttfamily 1405.3864}}].

\bibitem{Horowitz:2015dta}
W.~A. Horowitz, \emph{{Fluctuating heavy quark energy loss in a strongly
  coupled quark-gluon plasma}},
  \href{https://doi.org/10.1103/PhysRevD.91.085019}{\emph{Phys. Rev.}
  {\bfseries D91} (2015) 085019}
  [\href{https://arxiv.org/abs/1501.04693}{{\ttfamily 1501.04693}}].

\bibitem{Rajagopal:2016uip}
K.~Rajagopal, A.~V. Sadofyev and W.~van~der Schee, \emph{{Evolution of the jet
  opening angle distribution in holographic plasma}},
  \href{https://doi.org/10.1103/PhysRevLett.116.211603}{\emph{Phys. Rev. Lett.}
  {\bfseries 116} (2016) 211603}
  [\href{https://arxiv.org/abs/1602.04187}{{\ttfamily 1602.04187}}].

\end{thebibliography}\endgroup

\end{document}